\documentclass[fleqn,usenatbib]{mnras}

\usepackage{newtxtext,newtxmath}
\usepackage[T1]{fontenc}

\usepackage{tabularx}
\newcolumntype{R}[1]{>{\raggedleft\arraybackslash}p{#1}}

\DeclareRobustCommand{\VAN}[3]{#2}
\let\VANthebibliography\thebibliography
\def\thebibliography{\DeclareRobustCommand{\VAN}[3]{##3}\VANthebibliography}

\newcommand{\jname}{J173438.35$-$504550.4}
\newcommand{\mdmunits}{{\rm pc \, cm^{-3}}} 
\newcommand{\dmunits}{$\mdmunits$}
\newcommand{\mdmfrb}{{\rm DM}_{\rm FRB}}
\newcommand{\dmfrb}{$\mdmfrb$}
\newcommand{\mdmism}{{\rm DM}_{\rm ISM}}

\newcommand{\mdmhalo}{{\rm DM}_{\rm Halo}}

\newcommand{\mdmcosmic}{{\rm DM}_{\rm cosmic}}
\newcommand{\dmcosmic}{$\mdmcosmic$}
\newcommand{\mdmhost}{{\rm DM}_{\rm Host}}
\newcommand{\dmhost}{$\mdmhost$}
\def\arcsec{\hbox{$^{\prime\prime}$}}

\newcommand{\frbA}{FRB~20200413A}
\newcommand{\frbB}{FRB~20200915A}
\newcommand{\frbC}{FRB~20201123A}

\newcommand{\oii}{[O~{\sc ii}]}
\newcommand{\oiii}{[O~{\sc iii}]}
\newcommand{\nii}{[N~{\sc ii}]}
\newcommand{\ha}{H$\alpha$}
\newcommand{\hb}{H$\beta$}

\newcommand{\mstar}{11.2} % log10 Mstar
\newcommand{\sfr}{0.2}    %  SFR (Msun/yr)

\newcommand{\pybdsf}{{\tt pyBDSF}}

\usepackage{graphicx}	% Including figure files
\usepackage{amsmath}	% Advanced maths commands
\title[MeerTRAP FRB discoveries and localisations]{First discoveries and localisations of Fast Radio Bursts with MeerTRAP: a real-time, commensal MeerKAT survey}

\author[K. M. Rajwade et al.]{K. M. Rajwade,$^{1,2}$\thanks{E-mail: rajwade@astron.nl}
M. C. Bezuidenhout,$^{1}$
M. Caleb,$^{1,3,4}$
L. N. Driessen,$^{5}$
F. Jankowski,$^{1}$
M. Malenta,$^{1}$
\newauthor
V. Morello,$^{1}$
S. Sanidas,$^{1}$
B. W. Stappers,$^{1}$
M. P. Surnis,$^{1}$
E. D. Barr, $^{6}$
W. Chen, $^{6}$
M. Kramer, $^{6,1}$
J. Wu, $^{6}$
\newauthor
S. Buchner,$^{7}$
M. Serylak,$^{8,19}$
F. Combes,$^{9}$
W. Fong,$^{10}$
N. Gupta,$^{11}$
P. Jagannathan,$^{12}$
C. D. Kilpatrick, $^{10}$
\newauthor
J.-K. Krogager,$^{13}$
P. Noterdaeme,$^{14,15}$
C. N\'{u}n\~{e}z, $^{16}$
J. Xavier Prochaska, $^{17,18}$
R. Srianand,$^{11}$
and N. Tejos $^{16}$
\\
$^{1}$Jodrell Bank Centre for Astrophysics, University of Manchester, Oxford Road, Manchester M13 9PL, UK\\
$^{2}$ASTRON, the Netherlands Institute for Radio Astronomy, Oude Hoogeveensedijk 4, 7991 PD Dwingeloo, The Netherlands\\
$^{3}$Sydney Institute for Astronomy, School of Physics, The University of Sydney, NSW 2006, Australia\\
$^{4}$ASTRO3D: ARC Centre of Excellence for All-sky Astrophysics in 3D, Canberra 2601, ACT, Australia\\
$^{5}$ CSIRO, Space and Astronomy, PO Box 1130, Bentley, WA 6102, Australia\\
$^{6}$Max-Planck-Institut f\"ur Radioastronomie, Auf dem H\"ugel 69, D-53121 Bonn, Germany \\
$^{7}$ South African Radio Astronomy Observatory, Black River Park, 2 Fir Street, Observatory, Cape Town, 7925, South Africa\\
$^{8}$ SKA Observatory, Jodrell Bank, Lower Withington, Macclesfield, SK11 9FT, United Kingdom\\
$^{9}$ Observatoire de Paris, LERMA, Coll\`ege de France, CNRS, PSL Univ., Sorbonne Univ., 75014 Paris, France\\
$^{10}$Center for Interdisciplinary Exploration and Research in Astrophysics (CIERA) and Department of Physics and Astronomy, Northwestern University, Evanston, IL 60208, USA\\
$^{11}$ Inter-University Centre for Astronomy and Astrophysics, Post Bag 4, Ganeshkhind, Pune 411 007, India \\
$^{12}$ National Radio Astronomy Observatory, 1003 Lopezville Road, Socorro, NM 87801, USA\\
$^{13}$ Department of Astronomy, University of Geneva, Chemin Pegasi 51, 1290 Versoix, Switzerland\\
$^{14}$ Institut d'Astrophysique de Paris, 98 bis boulevard, Arago,75014, Paris, France\\
$^{15}$ Franco-Chilean Laboratory for Astronomy, IRL 3386, CNRS and Departamento de Astronom\'ia, Universidad de Chile, Casilla 36-D, Santiago, Chile\\
$^{16}$ Instituto de F\'isica, Pontificia Universidad Cat\'olica de Valpara\'iso, Casilla 4059, Valpara\'iso, Chile\\
$^{17}$ University of California, Santa Cruz, 1156 High St., Santa Cruz, CA 95064, USA\\
$^{18}$ Kavli Institute for the Physics and Mathematics of the Universe,
5-1-5 Kashiwanoha, Kashiwa, 277-8583, Japan \\
$^{19}$ Department of Physics and Astronomy, University of the Western Cape, Bellville, Cape Town, 7535, South Africa\\
}

\date{Accepted XXX. Received YYY; in original form ZZZ}

\pubyear{2022}

\begin{document}
\label{firstpage}
\pagerange{\pageref{firstpage}--\pageref{lastpage}}
\maketitle

\begin{abstract}
We report on the discovery and localization of fast radio bursts (FRBs) from the MeerTRAP project, a commensal fast radio transient-detection programme at MeerKAT in South Africa.  Our hybrid approach combines a coherent search with an average field-of-view of 0.4\,$\rm deg^{2}$ with an incoherent search utilizing a field-of-view of $\sim$1.27\,$\rm deg^{2}$ (both at 1284~MHz). Here, we present results on the first three FRBs:
\frbA\ (DM=1990.05~pc~cm$^{-3}$),
  \frbB\ (DM=740.65~pc~cm$^{-3}$),
  and
  \frbC\ (DM=433.55~pc~cm$^{-3}$).
\frbA\ was discovered only in the incoherent beam.
\frbB\ (also discovered only in the incoherent beam) shows speckled emission in the dynamic spectrum which cannot be explained by interstellar scintillation in our Galaxy or plasma lensing, and might be intrinsic to the source. \frbC\ shows a faint post-cursor burst about 200~ms after the main burst and warrants further follow-up to confirm whether it is a repeating FRB.
\frbC\ also exhibits significant temporal broadening consistent with scattering by
a turbulent medium.
The broadening exceeds that predicted for medium along the sightline
through our Galaxy. We associate this scattering with the turbulent medium in the environment of the FRB in the host galaxy.  
Within the approximately $1'$ localization region of \frbC\ ,
we identify one luminous galaxy ($r \approx 15.67$; \jname)
that dominates the posterior probability for a host association.
The galaxy's measured properties are consistent with other FRB hosts with secure associations.
% Recent all-sky radio transient searches have enabled rapid discovery, follow-up and localisations thus, elucidating some aspects of Fast Radio Bursts (FRBs) and emphasizing the need for new discoveries and localisations of FRBs. Here, we present the first three FRB discoveries from the MeerTRAP project, 
%a commensal FRB-detection programme at the MeerKAT interferometer in South Africa. 
%FRB~20200413A is the very first FRB observed only in the incoherent beam. It exhibits broadband emission with a hint of scattering at lower frequencies. FRB~20200915A shows speckled emission in the dynamic spectrum which cannot be explained by interstellar scintillation in our Galaxy or plasma lensing, and might be intrinsic to the source. FRB~20201123A shows evidence for scattering, along with a fainter burst (post-cursor) about 200~ms after the bright burst. The scattering in FRB~20201123A cannot be explained by our Galaxy and we show that it most likely originates from the vicinity of the burst. We also were able to get stringent constraints on the location of FRB~20201123A and report on a potential host galaxy from further optical follow-up.
\end{abstract}

\begin{keywords}
stars:neutron -- radio continuum:transients
\end{keywords}

%%%%%%%%%%%%%%%%%%%%%%%%%%%%%%%%%%%%%%%%%%%%%%%%%%

%%%%%%%%%%%%%%%%% BODY OF PAPER %%%%%%%%%%%%%%%%%%

\section{Introduction}
FRBs are bright radio flashes of hitherto unknown origin. They last for less than a few milliseconds and their dispersion measures (DMs), the integrated electron densities along the lines of sight, far exceed the contributions from our own Galaxy, indicating their cosmological nature~\citep{lorimer2007L, thornton2013}. FRBs are therefore potentially new probes to study the cosmic history of the Universe and are currently in use to probe important cosmological milestones~\citep[e.g.][]{macquart2020}. Since their discovery in 2007~\citep{lorimer2007L}, more than 600 FRBs  have been reported~\citep{chimecatalog}. Initially, most FRBs were observed as one-off events i.e., a single burst detected from a given part of the sky. The lack of multiple pulses from the same FRB suggested cataclysmic models to explain their origins~\citep[see][for a full review of theories]{Platts2019}. This changed when a FRB was found to repeat~\citep{spitler2016}. This was followed by its localisation to a host galaxy confirming their extragalactic nature. Moreover, evidence of a periodicity in the activity cycle of two repeating FRBs~\citep{Rajwade2020,r3periodic} combined with a detection of a FRB-like radio pulse from a highly magnetized neutron star located in our Galaxy~\citep{SGR1935chime,bochenek2020} were major breakthroughs towards constraining the progenitors of these enigmatic bursts. In the last few years, the field has progressed rapidly, owing to a slew of new repeaters discovered by the Canadian Hydrogen Intensity Mapping Experiment (CHIME), and the localisations and host galaxy identifications of many one-off FRBs discovered by the Australian SKA Pathfinder (ASKAP) telescope~\citep{shannon2018}. While the rate of FRB discoveries has increased tremendously, every new FRB shows interesting emission characteristics and morphology that pose more questions. For example, there is an obvious dichotomy between the spectral features and pulse widths of repeating and one-off FRBs~\citep{pleunis2021} but their rates are still consistent with all FRBs originating from a single underlying population~\citep{caleb2019}. This highlights the importance of discovering and following up these sources
at radio, optical, and other frequencies.

Radio transient surveys have expedited progress in the field of transient astrophysics in the recent years. The use of state of the art Central Processing Units (CPUs) and Graphical Processing Units (GPUs) can enable processing of high volumes of telescope data in real-time. Furthermore, with the advent of new technology, astronomers have been able to increase the field-of-view (FoV) and sensitivity of radio telescopes.  All of this has led to commensal transients searches where by the radio data taken for a different science goal are being used to look for FRBs and other radio transients. This is an important approach, especially when telescope resources are oversubscribed and a limited amount of time is available to do achieve all the scientific goals. Additionally, the need to not only discover new FRBs but the need to localise them and identify their host galaxies is paramount to elucidate some of these mysteries surrounding the FRB phenomenon. The MeerKAT radio telescope~\citep{jonas2016} in South Africa is the ideal telescope to perform this task by the virtue of its high sensitivity to transients and excellent angular resolution to localise. MeerTRAP (\textit{More} TRansients And Pulsars) is an ERC funded project (PI: Stappers) and has been deployed on the telescope to commensally discover and localise FRBs and other transients in real-time.

MeerTRAP piggybacks on other Large Survey Projects (LSPs) that are using MeerKAT for science observations.
During these commensal observations, MeerTRAP performs time domain searches for transients in real-time with the primary goal of discovering and localising FRBs in order identify their host galaxies. The instrument was commissioned in early 2019 and has been fully operational on all LSPs at MeerKAT since September of 2020. Since then, MeerTRAP has been discovering new radio transients within and beyond the Galaxy~\citep{bezuidenhout2022}. In this paper, we present the first three FRB discoveries by the MeerTRAP instrument. The paper is organized as follows: In Section~\ref{sec:mtp}, we give a brief overview of the MeerTRAP project. Then we present the new FRBs discovered and discuss some interesting emission properties for a few of them. In the next section, we discuss the localisation constraints on the FRBs, specifically focusing on one of them and discuss its potential host galaxy identification. Finally, we present our discussion and summarize our results in Sections~\ref{s:discussion} and \ref{s:summary} respectively.

\section{The MeerTRAP System}
\label{sec:mtp}
The MeerTRAP compute cluster also known as the Transient User Supplied Equipment (TUSE) consists of 67 servers with one head node and 66 compute nodes located in the Karoo Array Processing Building at the MeerKAT site. Each compute node contains two Intel Xeon 8C/16T CPUs, each possessing 16 logical cores for computation, two Nvidia GeForce 1080 Ti GPUs and 256 GB of DDR4 Random Access Memory (RAM) blocks. Each node is connected to a breakout switch via 10 GbE network interface cards (NIC) that are used to ingest data. 

The signals from all the antennas are detected and summed to form an incoherent beam (IB) that samples the entire field of view (FoV) of the telescope ($\sim$ 1.27~$\rm deg^{2}$ at 1284~MHz). Simultaneously, signals from a subset of antennas are added by using the phase information to create a large set of narrow, highly sensitive coherent beams (CBs) that sample a fraction of the primary FoV. Typically, the coherent beams are more sensitive than the incoherent beam by a factor of $\sim$5 but cover only a fraction of the primary FoV of the telescope ($\sim$0.4~deg$^{2}$ at 1284~MHz). However, this fraction depends strongly on the observing frequency and elevation as the telescope array projected on the sky modifies the shape of the CBs significantly~\citep[see][for a detailed treatment]{chen2021}. The coherent beams' positions are determined by an algorithm that efficiently arranges them within a circular tiling region such that they intersect at a user-specified relative sensitivity. By default MeerTRAP specifies that the CBs intersect at 25~per~cent of their maximum sensitivity. The full details of the method used for modelling the CB point-spread function are provided in~\cite{chen2021} and further details about the full algorithm  along with the verification and validation tests are being compiled as a separate paper (Bezuidenhout et al, submitted.). The incoherent and coherent beams are created in the Filterbank BeamFormer User Supplied Equipment (FBFUSE) cluster that has been built by the Max Planck Institute for Radio Astronomy in Bonn~\citep[see][for more details]{barr2018}. Commissioning tests in early 2019 have shown that the beamforming efficiency at 1284~MHz is typically between 0.92--0.96 which gave us confidence that we were able to exploit almost the entire sensitivity of the array. The results of this commissioning are presented in~\cite{chen2021}.
Data from FBFUSE are received over the network on the NICs as SPEAD2\footnote{\url{https://casper.ssl.berkeley.edu/wiki/SPEAD}} packets that are read by the data ingest code and written to POSIX shared memory ring buffers of 50 seconds duration. The data are arranged such that each compute node ingests a number of coherent beams to be processed in real-time. Since the data from the beamformer arrive in a frequency-time format (i.e. frequency being the slowest moving axis), they are transposed to a time-frequency format on a per beam basis, as required by the transient search code. The data are also scrunched in time and frequency, resulting in an effective time resolution of 306~$\mu$s over a band of 856~MHz that is split into 1024 channels (channel width of 208.4~kHz), 4096 channels (channel width of 104.2~kHz) depending on the number of frequency channels that are scrunched in the beamformer. The resulting filterbank data are saved in separate shared memory buffers corresponding to each beam. Details of the MeerTRAP instrumentation have been presented in \citet{2018sanidas, csa+20}, \citet{2020jankowski}, \citet{2020malenta}, \citet{Rajwade2020}, and a complete system overview will be given in Stappers et al.\ (in prep.).

\begin{table}
% \centering
\caption{Dedispersion plan for single pulse search pipeline.}
\label{tab:search}
% \vspace*{5mm}
\centering
\begin{tabular}{r l l}
\hline
DM range & $\Delta$ DM & Downsampling factor\\ [1ex]
(pc~cm$^{-3}$) & (pc~cm$^{-3}$) &  \\
\hline
 0.00 -- 383.75  & 0.307 & 1 \\
 383.75 -- 774.95 & 0.652 & 2 \\
 774.95 -- 1534.55 & 1.266  & 4 \\
 1534.55 -- 3041.75 & 2.512 & 8 \\
 3041.75 -- 5241.75  & 4.000  & 16  \\
\hline
\end{tabular}
\end{table}

The data at 1.284~GHz for each beam are searched for bright bursts using the state-of-the-art, GPU-based single pulse search pipeline \textsc{AstroAccelerate}\footnote{\url{https://github.com/AstroAccelerateOrg/astro-accelerate}} \citep{armour2011gpu, adamek2020single}. The real-time search was done by incoherently de-dispersing in the DM range 0--5118.4~pc cm$^{-3}$, divided into multiple sub-ranges with varying DM steps and time averaging factors (see Table~\ref{tab:search} for details) across a bandwidth of 856~MHz with a sampling interval of 306~$\mu$s. We also searched up to a maximum boxcar width of 0.67 s. This particular choice of parameters allowed us to process all data in real time, thanks to strict optimisations applied in the \textsc{AstroAccelerate} algorithms. 

To reduce the number of detections due to Radio Frequency Interference (RFI), we applied a static frequency channel mask to the data before the de-dispersion and single-pulse search. While the RFI remained stable most of the time, the static mask did suffer from instances where the channels were corrupted by RFI varied, resulting in many spurious detections.

%In late summer of 2020, we deployed a new RFI mitigation algorithm called Inter-Quartile Range Mitigation (IQRM) that using the standard deviation of the frequency band to detect spurious signals and create a time variable mask for the data. More details of the new algorithm are presented in~\cite{Morello2021}. IQRM resulted in a significant reduction in the number of false positives and also removed the need to use a static channel thus, giving us a larger effective bandwidth without the increase in RFI candidates. On average, about 25$\%$ of the channels are masked by IQRM at 1.4~GHz.
Additionally, the data are filtered using standard zero-DM excision \citep{eatough2009} to remove any remaining broadband RFI that was infrequent enough not to be included in the mask. Then, the candidates are sifted based on some static cutoffs of DMs below 20~pc~cm$^{-3}$, widths above 300~ms and signal-to-noise (S/N) below 8.0 to reduce the number of noise or RFI candidates. The resulting candidates are then clustered using the Friends-of-Friends algorithm~\citep[see][for more details]{huchra1982} by combining candidates that are in close proximity in time and DM space. This resulted in a significant reduction of the total number of output candidates ($\sim$ 40$\%$). The final list of candidates are then saved to disk for further inspection.The extracted candidate files contain raw filterbank data of the dispersed pulse and additional padding of 0.5~s at the start and at the end of the file. This entire pipeline with all the aforementioned steps is encapsulated in a modular \textsc{C++} framework called \textsc{cheetah}~\footnote{\url{https://gitlab.com/SKA-TDT/cheetah}} which is deployed on the compute nodes of MeerTRAP.

\section{FRB discoveries}

\begin{table*}
% \centering
\caption{Measured quantities for the first three FRBs discovered by MeerTRAP. Note that for \frbA~and \frbB, the Right Ascension (RA) and Declination (Dec) are coordinates of the boresight of the observation and not the true position. The DMs listed here are the S/N maximized values which are different from the value at which the burst was detected. The values in parenthesis denote the formal errors on the most significant digits.}
\label{tab:specparams}
% \vspace*{5mm}
\centering
\begin{tabular}{l l l l l l l l l l l l}
\hline 
Name  & Barycentric MJD & RA& Dec  & gl & gb & S/N & DM & Width & DM$_{\rm ne2001}$ & DM$_{\rm ymw16}$ & $\mathcal{F}$ \\ [1ex]
 & & ($^\circ$) & ($^\circ$) & ($^\circ$) & ($^\circ$) &  & (pc~cm$^{-3}$) & (ms) & (pc~cm$^{-3}$) & (pc~cm$^{-3}$) & (Jy~ms) \\ [1ex]
\hline
FRB~20200413A & 58952.382356 & 328.57 & $-$28.21 & 20.72 & $-$50.98 & 14 & 1990.05 (88) & 4.9 (2) & 39.2 & 26.22 & $\gtrsim$3.1 \\
FRB~20200915A & 59107.105051 & 42.41  & 4.67 & 169.35 & $-$47.25 & 35 & 740.65 (40) & 2.2 (1) & 39.1 & 33.72 & $\gtrsim$3.0 \\
FRB~20201123A & 59176.421148 & 263.67 & $-$50.76 & 340.229 & $-$9.681 & 45 & 433.55 (36) & 4.6 (2)&  251.93 & 162.4 & $\gtrsim$1.4\\
\hline
\end{tabular}
\label{tab:params}
\end{table*}

\begin{figure*}
 \centering
\includegraphics[scale=0.42]{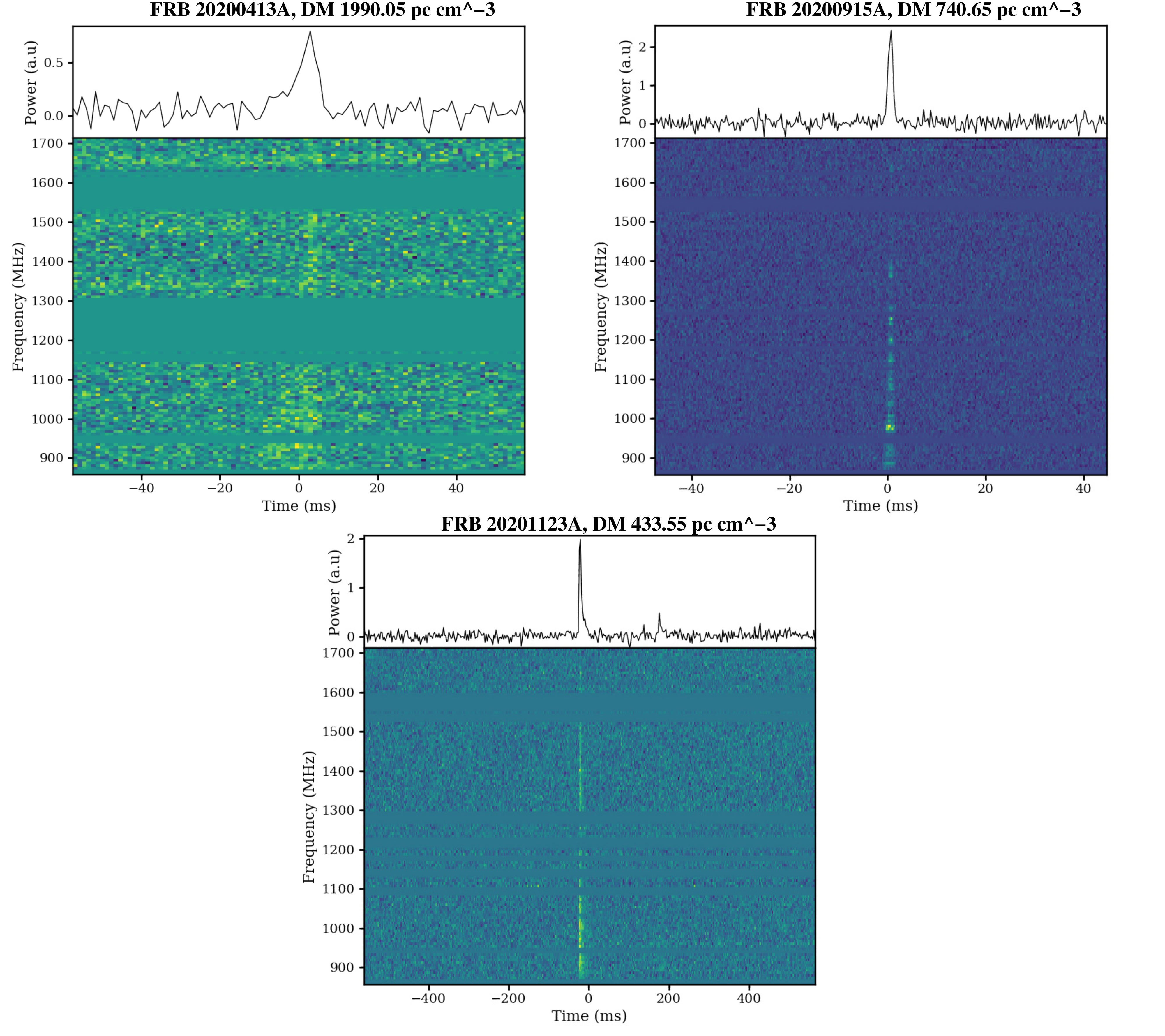}

\caption{Dynamic spectrum of all the FRBs presented in this paper. The lower panels show the time frequency data while the top panel shows the frequency integrated burst profile. The pulses have been dedispersed to the DM that maximizes the S/N of the detection. The horizontal lines that show the same colour are data that have been flagged due to RFI.}
\label{fig:dyn} 
\end{figure*} 

 The dynamic spectra of the first three FRB discoveries are shown in Figure~\ref{fig:dyn} and the measured and derived parameters for them are presented in table~\ref{tab:params}. We note here that beam size and shape of the CB and the IB strongly depend on the frequency and the elevation of the source. This is an important factor contributing to the FRBs being brighter in the lower half of the band and may not represent the true emission dependence on observing frequency. The three FRBs reported here are all discovered far off the Galactic Plane (gb: $-$10$^\circ$ -- $-$50$^\circ$).

\subsection{FRB~20200413A}

This is the first FRB discovered by MeerTRAP. It was detected in the Incoherent Beam (IB) on UT 13$^{\rm th}$ April 2020 during commensal observations of the pulsar timing experiment at MeerKAT \citep[MeerTime;][]{meertime}. The FRB was only detected in the IB with no detections in the CB suggesting that the FRB was outside the more sensitive CB tiling pattern. The source was detected in the real-time
pipeline with a S/N of 9.1 and a DM of 1988~pc~cm$^{-3}$. The S/N and the DM were then optimized to values of 14 and 1990.05~pc~cm$^{-3}$ respectively using a custom made
S/N-DM optimization code \textsc{mtcutils}\footnote{\url{https://bitbucket.org/vmorello/mtcutils/}}. The
FRB spans the entire  bandwidth of the MeerKAT L-Band receiver (856--1712~MHz) and shows broadening of the pulse width at lower
frequencies probably due to the increase in the intra-channel dispersion smearing (t$_{\rm DMsmear}$
$\simeq$ 20~ms at a frequency of 856~MHz for 1024 channels) and scattering. Unfortunately, the S/N of the burst is too low for
a thorough analysis of this smearing.

\subsection{FRB~20200915A}

FRB~20200915A was discovered on UT 15$^{\rm th}$ of September 2020, again only in the IB. This FRB is much
brighter than FRB~20200413A and was detected with a S/N of 29. Subsequent optimization of the S/N and DM
with \textsc{mtcutils} showed that the FRB has S/N ratio of 35 at the best DM of 740.65~pc~cm$^{-3}$. The FRB was discovered when MeerTRAP was commensally observing with the MeerKAT Absorption Line Survey \citep[MALS;][]{Gupta2016mals} observation. The lack of CB detections meant that localising the FRB based solely on the IB
detection was not constraining. FRB~20200915A shows broadening
at lower frequencies along with evidence of scintillation in the dynamic
spectrum akin to what is seen in pulses from radio emitting neutron stars in our Galaxy~\citep[for e.g.][]{Cordes98}.

\subsection{FRB~20201123A}

FRB~20201123A was discovered on UT 23$\rm ^{rd}$ of November, 2020. It was detected in a single CB with no other detections in any other neighbouring CBs nor in the IB, making this the first discovery in the coherent beams. The burst was discovered at a S/N of 30.6 during a commensal observing run with the MeerTIME project~\citep{meertime}. The best optimized DM with \textsc{mtcutils} is 433.55~pc~cm$^{-3}$ with a S/N of 45. The burst shows hints of scatter broadening in the bottom half of the band. Interestingly, there is a faint post-cursor seen in the time series separated from the main burst by $\sim$200~ms. The burst separation is consistent with separations observed for bursts from other repeating FRBs like FRB~20121102A~\citep{Cruces2021} suggesting that \frbC~could be a repeater.

\section{Scattering and Scintillation}

% Scintillation ACF for FRB 200915
\begin{figure}
 \centering
\includegraphics[width=3.5 in]{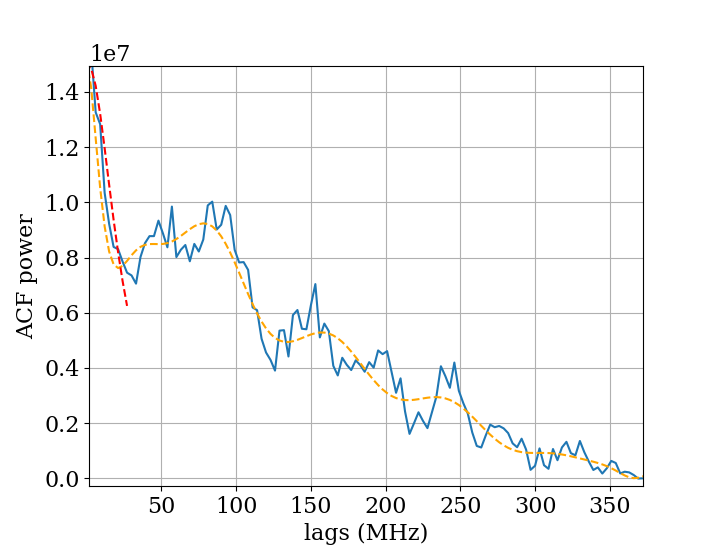}
\caption{Autocorrelation function for FRB~20200915A as a function of lag in MHz. The red dashed curve shows the best fit for Lorentzian function to the very first peak after ignoring the zero-lag component that is attributed to self-noise. The orange curve shows a polynomial fit to the rest of the ACF to show the general trend of the dynamic spectrum to aid the viewer.}
\label{fig:acf} 
\end{figure} 

\subsection{FRB~20200915A}

While FRB~20200915A does not show any obvious signatures for interstellar scattering, the dynamic spectrum clearly shows islands of band-limited emission in the lower half of the bandwidth (Figure~\ref{fig:dyn}). While this could be intrinsic to the source, such emission behaviour is typically an indication of diffractive interstellar scintillation (DISS)~\citep{Rickett70}. In order to quantify whether this is indeed scintillation, we computed the scintillation bandwidth. To do that, we selected the dynamic spectrum (after removal of channels corrupted by RFI) for only the time bins corresponding to the on-pulse region of the burst. Then the data were added along time axis to generate a frequency spectrum that corresponds just to the pulse integrated to a single time bin. Since the emission is dominated by the lower-half of the frequency band, we decided to perform the analysis on just the lower half (856--1284~MHz) of the band. Then, we computed the discrete auto-correlation function (ACF) of this spectrum. For a given signal $S(\nu)$,
\begin{equation}
    \text{ACF}(\Delta\nu) = \sum_{\nu = 1}^{\rm nchan} S(\nu)\overline{S(\nu-\Delta\nu)}
\end{equation}
where $\Delta\nu$ is the lag in frequency axis, $S(\nu - \Delta\nu)$ is the signal at an observing frequency of $(\nu - \Delta\nu)$ and $\rm nchan$ is the total number frequency channels. The resulting ACF is shown in figure~\ref{fig:acf}. The decorrelation bandwidth of interstellar scintillation in the strong scattering regime is given by
the full-width at half-maximum (FWHM) of a Lorentzian function fit to the ACF~\citep{Cordes98}. After removing the DC component that can be attributed to
self-noise of the burst itself, we fit a Lorentzian function to the ACF to obtain the $\Delta\nu_{\rm DISS} \sim$14.75~MHz. To compute the expected scintillation bandwidth due to our own Galaxy along the line of sight to this FRB, we use the \textsc{ne2001} model for Galactic electron density~\citep{ne2001}. The \textsc{ne2001} model predicts the expected free electron density in the Galaxy based on estimates of DMs along various lines-of-sight from known radio pulsars. The predicted scintillation bandwidth is $\sim$3.4~MHz, a factor of $\sim$4 smaller than what is measured. We note that it is hard to gauge the significance of this deviation as no quantification of any formal model uncertainties is available currently in the literature. Hence, we can only say that assuming the model correctly predicts scintillation bandwidth, the deviation may be significant. Assuming that scintillation is caused by a thin screen with density fluctuations in the free electron density, one expects the scintles to become wider with frequency
($\Delta\nu_{\rm DISS} \sim \nu^{4}$ )~\citep[for e.g.][]{handbook}. The ACF of the dynamic spectrum of \frbB~along the frequency axis tells a different story. The islands of
emission become visually narrower with increasing frequency which suggests that the observed emission cannot be explained by the standard model of interstellar scintillation from our Galaxy~\citep{majid2021}. If we assume that the 14.75 MHz structure is due to scintillation from the Inter Galactic Medium (IGM) or halos of intersecting foreground galaxies~\citep{prochaska2019}, the corresponding scattering timescale, $\tau_{d}$ = $\frac{1}{2\pi~\delta \nu_{d}}$ = 10~ns which is almost four orders of magnitude smaller than our finest sampling interval. While it has been shown that the IGM may not be able to contribute meaningfully to the scintillation~\citep[see][and the references therein]{jp2013}, we cannot rule out that the structures we see are due to weak scintillation and scattering from foreground halos along the line of sight or the combination of the two~\citep{ravi2016}.

\begin{figure*}
 \centering
\includegraphics[width=3.5 in]{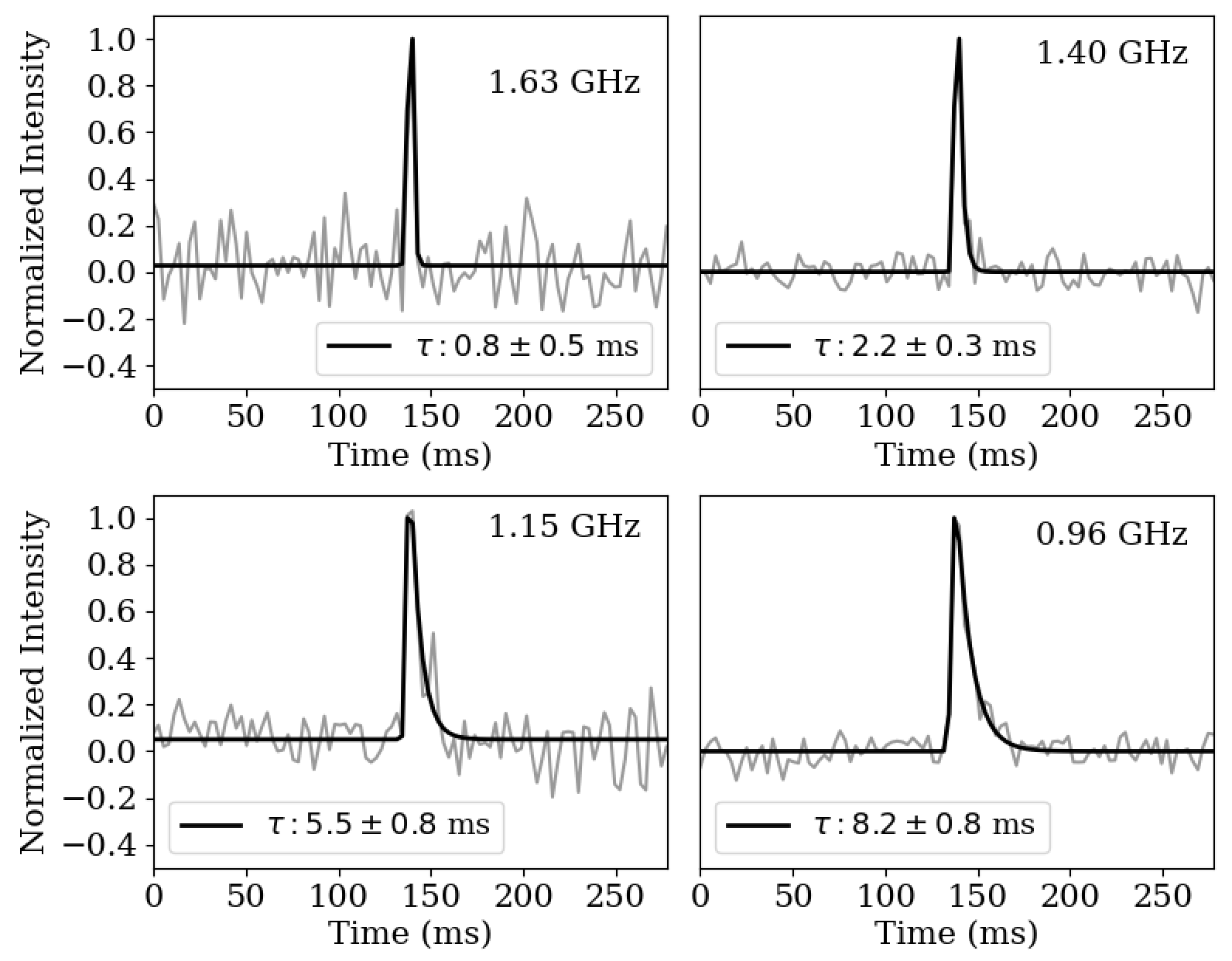}
\includegraphics[width=3.5 in]{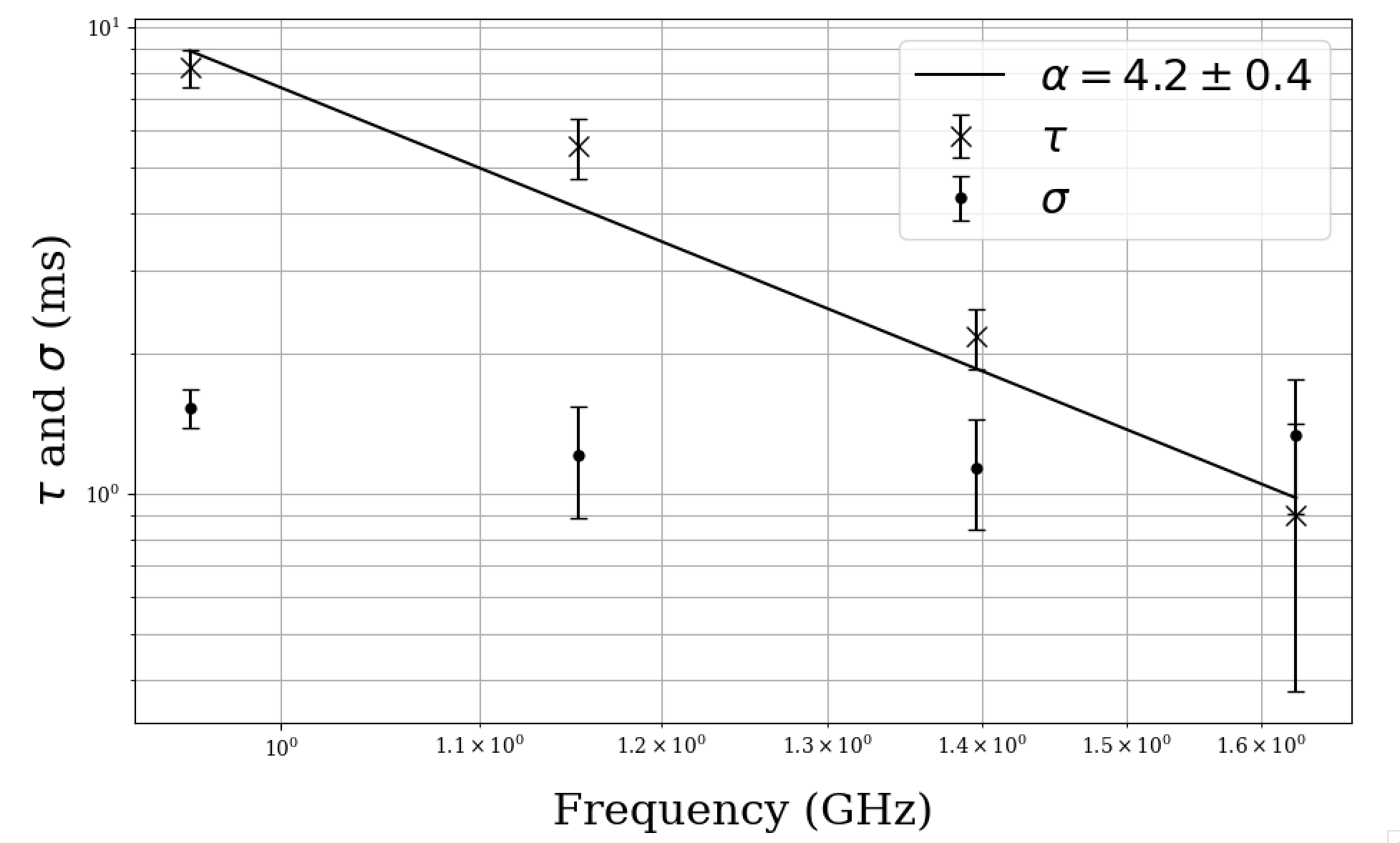}

\caption{\textbf{Left Panel:} FRB~20201123A in different subbands (grey line) along with the best model fit to the scatter broadening (black line). Each panel shows the frequency of the subband. \textbf{Right Panel:} The variation of the scattering timescale ($\tau$) as a function of observing frequency. $\alpha$ denotes the best-fit slope of the power-law function and the black line shows the best fit curve while $\sigma$ denotes the width of the Gaussian used to fit the burst at each frequency.}
\label{fig:scatter} 
\end{figure*} 

\subsection{FRB~20201123A}

The dynamic spectrum of FRB~20201123A shows evidence for scattering. To fully characterize the scattering, we divided the data into four subbands, each 214~MHz wide with a compromise between bandwidth and the S/N of the burst in each subband. We assume the scattering is caused by a thin scattering screen between the source and the observer that leads to an exponential tail in the resulting profile~\citep[see][and the references therein]{chawla2021}. Then, we used the \textsc{scamp-i} software suite~\citep{Oswald2021} to fit the burst profile at each frequency with an exponentially modified Gaussian~\footnote{\url{https://github.com/pulsarise/SCAMP-I}}. The code simultaneously fits for the scattering as well as the correction needed to the DM of the burst such that the peak of the burst under this model is aligned correctly in time after correcting for scatter broadening. The change in the DM from this fit is still consistent (within errors) with the optimized DM reported by \textsc{mtcutils} for this FRB (DM=433.55~pc~cm$^{-3}$). The fits are performed by sampling the likelihood function of the model using MCMC~\citep[see][ for more details]{Oswald2021}. Figure~\ref{fig:scatter} shows the result of the fits. The scattering timescale at the lowest subband is a factor of 2 more than the maximum expected smearing due to dispersion (4.7~ms at the lowest frequency), further supporting that the extended tail is due to scattering of the burst. The scattering timescale for this FRB seems to follow a powerlaw ($\tau \propto \nu^{-\alpha}$) with $\alpha$ = 4.2$\pm$0.4 which is consistent with what is expected from a scattering screen in our Galaxy based on the studies of known pulsars~\citep{qiu2020}. Despite that, the expected scattering at the best known Galactic latititude and longitude of FRB~20201123A (l=340.4, b=-9.67) cannot account for the scattering seen in this FRB~\citep{ne2001} which gives an expected $\tau$ of 0.016~ms at 1~GHz compared to the measured $\tau$ of 7.5~ms at 1~GHz. This points to a different source of scattering (the host galaxy or the IGM, including intersecting halos or the combination of the two) where the medium is more turbulent than what is observed in our Galaxy~\citep{chawla2021}. 

\section {Localisation}
For all these FRBs, we did not have visibility data correlations between different sets of antennas in the array) available hence localising the FRBs by imaging the data was not possible. Therefore, we rely on detections in the CBs and the IB to provide constraints on the location based on the best measured model of the CB and the IB~\citep{chen2021}. We also note that the error on the beam positions is much smaller than the beam width and so, negligibly contributes to the overall error on the position of the FRB. The complete methodology used and the corresponding validation and verification tests will be presented in an upcoming paper (Bezuidenhout et al., submitted.). The true position and the uncertainty of \frbA~and \frbB~are hard to gauge due the lack of detection in the CB and the FRBs are located anywhere between the FWHM of the IB (1.15$^\circ$ at 1284~MHz) and the edge of the CB tiling (as shown in detail below).
\begin{figure*}
 \centering
\includegraphics[scale=0.33]{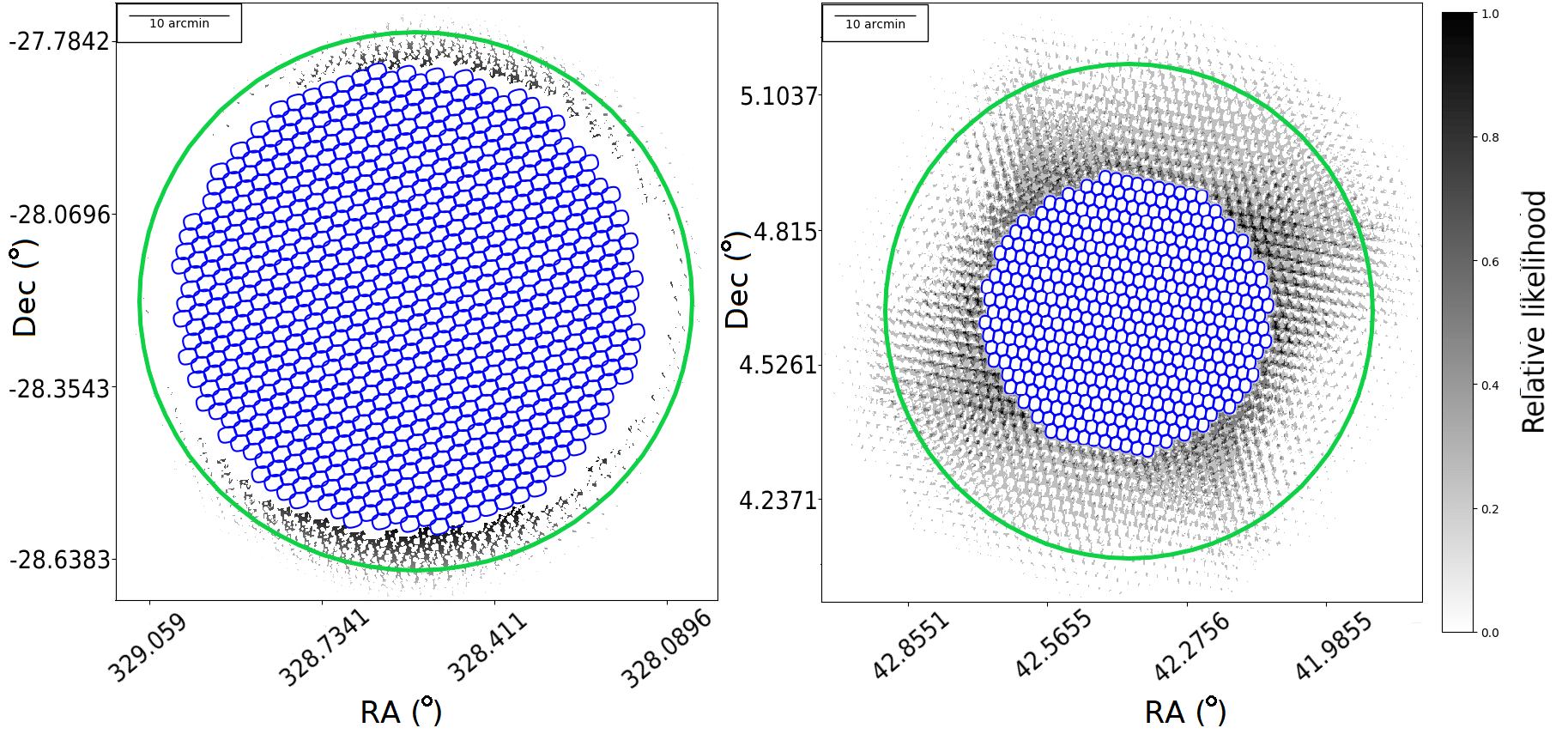}

\caption{Likelihood greyscale map of location of FRB~20200413A (left panel) and FRB~20200915A (right panel). The white central region marks the area with zero probability and corresponds to the location of the coherent beam tile (shown by the blue ellipsoids) where the FRBs cannot be located as they are only detected in the IB (see text for more details). The green circle marks the FWHM of the IB at 1284~MHz. The difference in the region covered by the CB tile depends on the total number of beams that are formed and the elevation of the source.}
\label{fig:frb12loc} 
\end{figure*}

\subsection{FRB~20200413A}

 A detection only in the IB meant that the localisation region for this FRB was unconstrained as the source could lie in any region where the IB was more sensitive than the CB (approximate CB FoV of 1.27~$\rm deg^{2}$). The exact localisation region is difficult to determine as beyond the primary beam, the beam response is highly asymmetric with multiple sidelobes with varying frequency dependence. However, the smooth appearance of 20200413A's observed dynamic spectrum over a wide band (see Figure~\ref{fig:dyn}) suggests that the source is not located in a far sidelobe of the IB, as the positions of the sidelobes are strongly frequency dependent. 

In order to more precisely constrain the region beyond the CB main lobes where the FRB might have originated, we used \texttt{Mosaic}~\citep{chen2021} to generate a PSF for the CBs at various frequencies, as well as sensitivity maps of the MeerKAT primary beam obtained using astrophysical holography \citep[e.g.][]{2021asad, devilliers2022}. The ratio of the sensitivity of a CB, S$_{\mathrm{CB}}$ to that of the IB, S$_{\mathrm{IB}}$, at a given point is given by
\begin{equation}
    \frac{S_{\mathrm{CB}}}{S_{\mathrm{IB}}} = \frac{N_{\mathrm{CB}}}{\sqrt{N_{\mathrm{IB}}}},
\end{equation}
where N$_{\rm CB}=36$ and N$_{\rm IB}=56$ are the number of antennas used to form the CB and IB during the observation, respectively.

At each frequency, and for each CB, all coordinates were excluded as a potential origin for the FRB where
\begin{equation}
    S/N_{\mathrm{CB}} > S/N_{\mathrm{IB}}\frac{N_{\mathrm{CB}}}{\sqrt{N_{\mathrm{IB}}}},
\end{equation}
where S/N$_{\mathrm{IB}}$ is the measured S/N in the IB at that frequency, and S/N$_{\mathrm{CB}}$ is the predicted CB S/N that frequency. Hence, all positions where any CB was more sensitive than the IB, are excluded. Viable positions were assigned a value of one and excluded positions were assigned a value of zero. 

This process was repeated at eight frequencies from 856~MHz to 1605~MHz, and the resulting maps added together are shown in the left-hand panel of Figure~\ref{fig:frb12loc}. The blue ellipsoids show the 25 per cent level\footnote{This number refers to the 25~per~cent level of the maximum sensitivity of the CBs at the L-band centre frequency of 1.284~GHz as determined using \texttt{Mosaic}.} of the main lobe of each CB. The colour scale corresponds to the number of sub-bands in which a given position was deemed viable; the maximum likelihood therefore occurs where the IB was more sensitive than all CBs in all sub-bands. This analysis indicates that the FRB most likely originated from immediately outside the CB tiling region.

\subsection{FRB~20200915A}

Similar to FRB~20200413A, FRB~20200915A has very little localisation information as it was only detected in the IB. The result of a similar localisation analysis as performed for FRB~20200413A is shown in the right-hand panel of Figure~\ref{fig:frb12loc}.

Since in this case MeerTRAP was piggybacking a MALS observation, the MeerKAT correlator saved correlations for every 8-second integration. This means that if the FRB is bright enough, it will be detected in the 8-second radio images of the field. For a pulse with a detected $\rm S/N_{\rm td}$ in the time domain search, the expected S/N in the image,
\begin{equation}
    {S/N_{\rm image}} = S/N_{\rm td} \frac{G_{\rm CB}}{G_{\rm IB}} \sqrt{\frac{W_{\rm td}}{T_{\rm img}}},
\end{equation}
where $G_{\rm CB}$ is the gain of the telescope when it is fully phased and the signals from all antennas are coherently added, $G_{\rm IB}$ is the gain of the incoherent sum, $W_{\rm td}$ the observed width of the FRB in the time-domain data and $T_{\rm img}$ is the integration time of the image. For a total of 60 dishes used in the observation we estimate $G_{\rm CB}\simeq$2.75~K~Jy$^{-1}$ and  $G_{\rm IB}\simeq$0.35~K~Jy$^{-1}$. We assume here that the beamforming efficiency was close to 1. Hence, for a S/N of 45 in the IB, we expect $S/N_{\rm image}$ of $\sim$ 4.4 in a 8-second integration radio image.

The MALS data were processed using the Automated Radio Telescope Imaging Pipeline (ARTIP). The details of ARTIP and data processing steps are provided in \citet[][]{gupta2021}. In short, we excluded edge frequency channels and applied a RFI mask to exclude the strong persistent RFI in the L-band. The data were then flagged and calibrated using the {\tt ARTIP-CAL} package.  The calibrated data were then processed using {\tt ARTIP-CONT} to perform wideband continuum imaging.  For this the calibrated target source data were averaged in frequency per 32 channels ($\sim$0.8\,MHz) and a more stringent RFI mask to completely exclude band edges and RFI-afflicted regions was applied.  Next, the resultant frequency-averaged 960 channels were regridded along the frequency axis to obtain a measurement set with 16 physically distinct spectral windows.  Three rounds of phase-only and one round of amplitude and phase self-calibration were performed.  For the widefield (6k $\times$6k image; pixel size = 2$^{\prime\prime}$) broadband imaging, the CASA task {\tt tclean} with {\tt w-projection} as the gridding algorithm, with 128 planes, in combination with {\tt Multi-scale Multi-term Multi-frequency synthesis} 
({\tt MTMFS}) for deconvolution, with nterms = 2 and four pixel scales to model the extended emission, were used.  
The images were deconvolved down to 3$\sigma$ using masks generated through the Python Blob Detector and Source Finder (\pybdsf\footnote{\href{https://www.astron.nl/citt/pybdsf/}{https://www.astron.nl/citt/pybdsf/}}). 
The final continuum image, made using {\tt robust=0} weighting, has a synthesized beam of  $9.2^{\prime\prime}\times6.6^{\prime\prime}$ (position angle = $-0.8^\circ$).  The continuum rms is $\sim$15\,$\mu$Jy\,beam$^{-1}$.

For the FRB localization, we used a self-calibrated dataset, to make broadband images for 23 timestamps within the time range: 02:23:03.8 - 02:25:59.8.  Since, the FRB has no signal above 1400\,MHz, we considered only spectral windows 0 to 10 covering 890 - 1415\,MHz.  These 6k $\times$6k timestamp images
\citep[pixel size = 2$^{\prime\prime}$; Briggs {\tt robust=0} weighting;][]{1995AAS...18711202b} at reference frequency of 1145.2\,MHz typically have resolution and rms of $12.7^{\prime\prime}\times7.5^{\prime\prime}$ and 160\,$\mu$Jy\,beam$^{-1}$, respectively.  The images were corrected for the primary beam attenuation using the model from the KATBEAM library\footnote{https://github.com/ska-sa/katbeam}. The resulting primary beam corrected images were used for the difference imaging in order to detect a transient source. For this purpose, we used two different methods: subtracting an average image and subtracting consecutive images. We performed both of these methods for both sets of 8\,second images. The average image was produced by adding all of the 23 8\,second images together and dividing by 23. For the consecutive difference images we simply subtracted the previous 8\,second image from the next image. After we had produced the sets of difference images, we used \pybdsf\ to extract sources from the images. We found that the signal to noise of the extracted sources in the difference images of the time step of the FRB, the time step before the FRB, and the timestep after the FRB was a normal distribution centered on S/N$\sim4$. A few S/N outliers (S/N$\gtrsim$7) were found, but were determined to be artefacts next to bright sources. As such we were unable to distinguish which source found by \pybdsf\,in the difference images may be the FRB.

\subsection{FRB~20201123A}

\begin{figure}
%\flushleft
\includegraphics[width=0.43\textwidth]{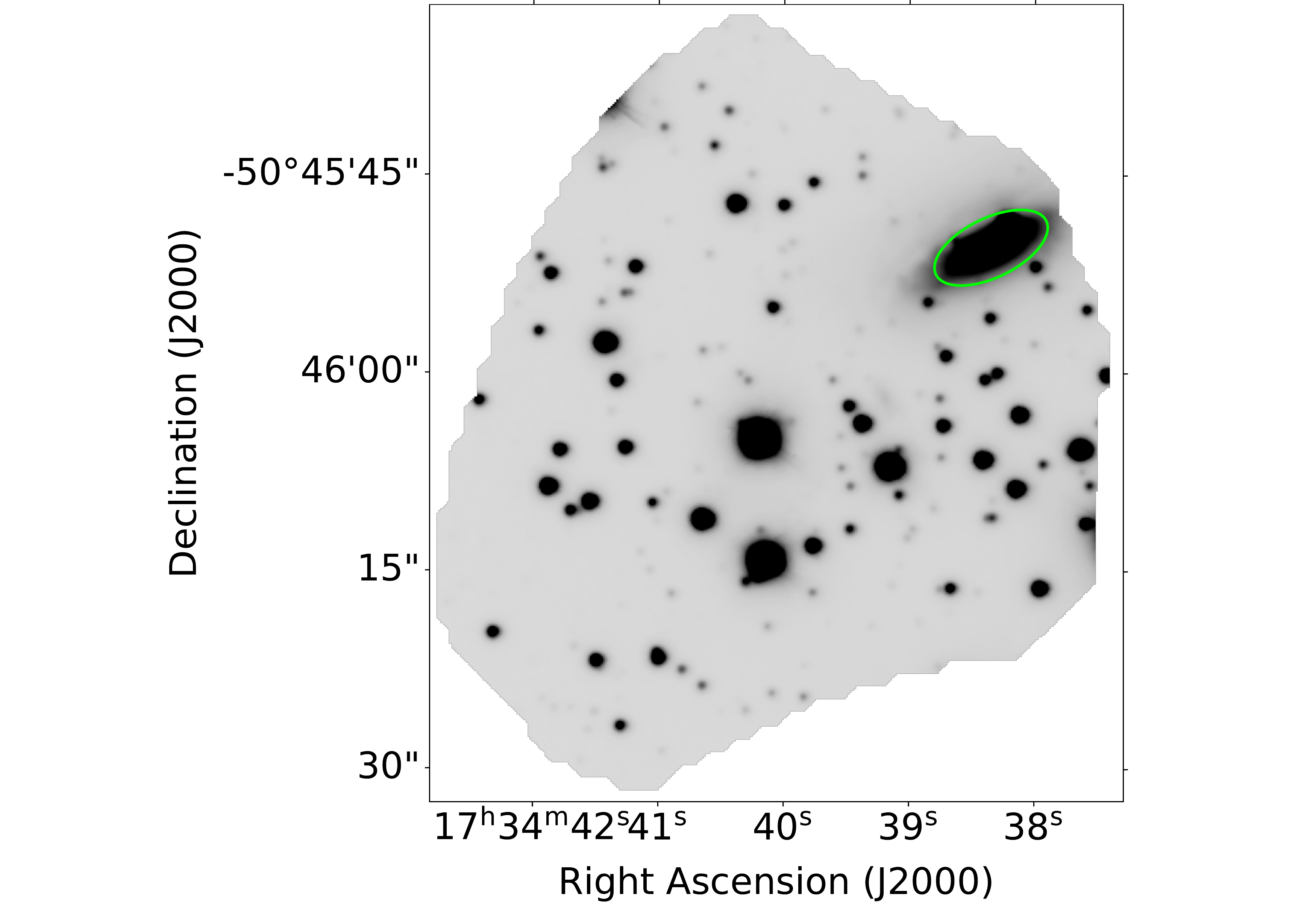}
\caption{GMOS-S $r$-band image of our localisation region
for FRB~20201123A. The figure shows GMOS-S data only within the 99$\%$ confidence region for the location of FRB~20201123A. The extended source \jname\ is 
the bright, most extended extragalactic source in the top right of the figure and
the leading candidate for the host of FRB~20201123A (green ellipse).
}
\label{fig:201123} 
\end{figure}

The detection of FRB~20201123A in only a single coherent beam (FWHM of $\sim$60 arc-seconds) does constrain its location to a 50"x51" ellipse within the coherent beam (see Figure~\ref{fig:201123}) using
\textsc{SeeKAT}\footnote{\url{https://github.com/BezuidenhoutMC/SeeKAT}}, a tied array beam localisation algorithm designed to constrain
location of bursts using detections in multiple beams (Bezuidenhout et al., submitted.). Hence, the uncertainties reported in
Table~\ref{tab:params} for \frbC~are 25" in RA and DEC. One can see that the localisation region is slightly smaller than the size of the CB at 25$\%$ power level and that is because of the added constraint of the lack of detections in adjacent CBs~\footnote{We note that the localisation region is strongly dependent on the beam spacing but for MeerTRAP, we have gone for a trade-off between the precision of localisation and the total sky covered by the CB tiling.}. The detection in a single coherent beam with no IB detection suggests that the
expected S/N in the IB was below the detection threshold of the search pipeline. While we did not save complex voltage data for this FRB, 
we are able to make several inferences on its potential host galaxy. In order to do that, we try and constrain the location using Bayesian inference on the burst within this single coherent beam. The main assumption we make here is that the intrinsic spectrum of the FRB is best characterized by a power-law function. In this scenario, for a burst with an intrinsic spectral index $\alpha$, the posterior probability of the offset from the boresight is

\begin{equation}
 %\begin{split}
    P(\phi,\alpha,S_{\rm t}| S_{\rm o}) \propto \mathcal{L}(\alpha, \phi, S_{\rm t}) 
    ~P(\alpha)P(\phi)P(S_{\rm t}(\nu_{0})),
 %\end{split}
\end{equation}
where the likelihood function is given by the following expression, having split the observing band into $n$ equal-sized sub-bands each with centre frequency $\nu_i$:
\begin{equation}
   \mathcal{L}(\alpha, \phi, S_{\rm t}) =  \prod_{i=0}^{n-1} {\rm exp}\left( -\frac{\left(S_{\rm o}(\nu_i) - G(\pmb{\phi}, \nu_i)S_{\rm t}(\nu_i) \right)^{2}}{2}\right),
\end{equation}

\begin{figure*}
%\flushleft
    \includegraphics[width=0.5\textwidth]{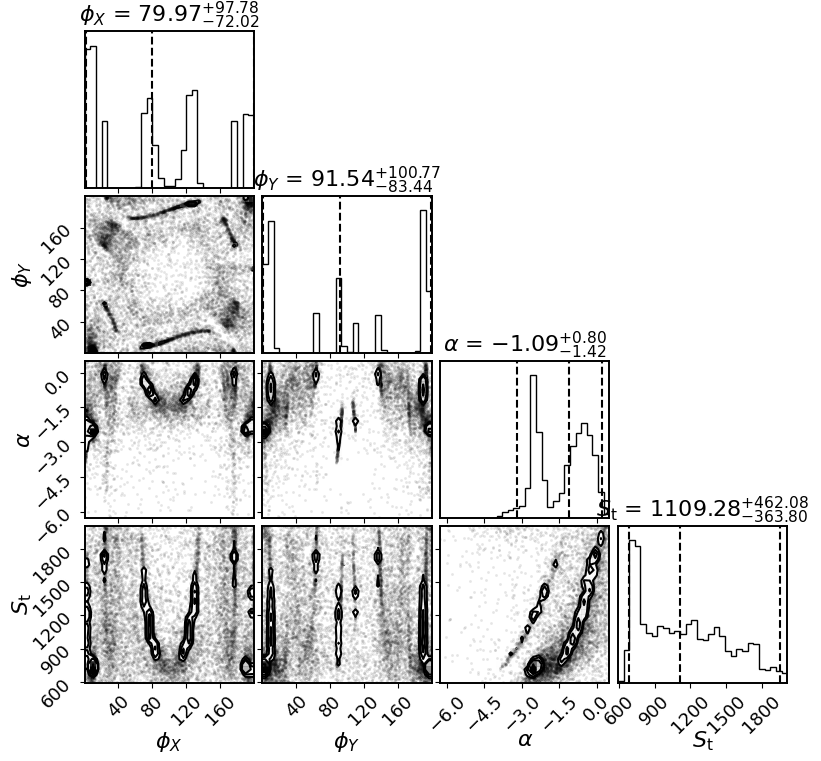}
     \includegraphics[width=0.49\textwidth]{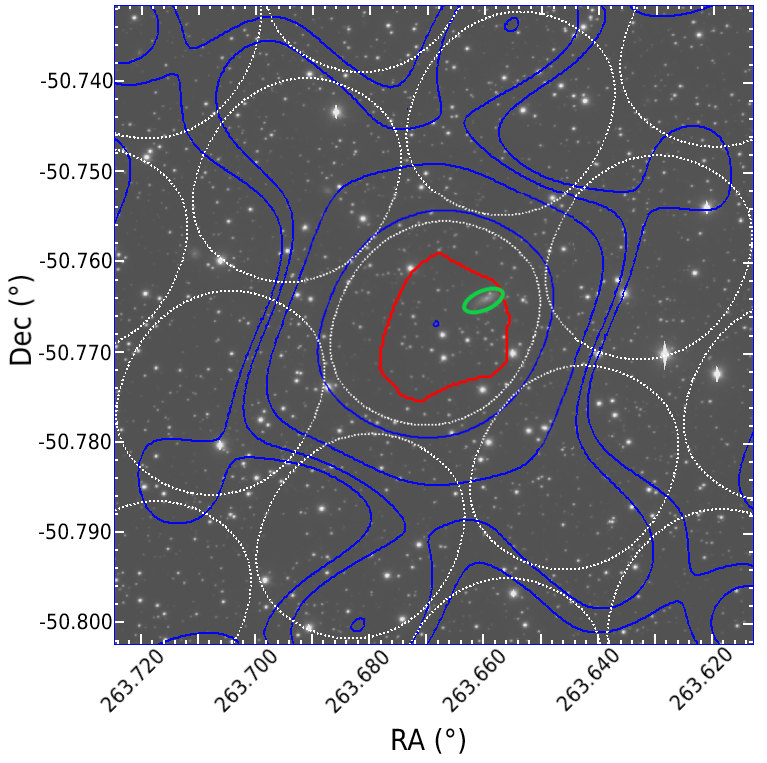}
\caption{\textbf{Left Panel}: Corner-Corner plot of the marginalized posterior distribution of the location of FRB~20201123A offset from the boresight ($\phi_{X}$, $\phi_{Y}$), the spectral index ($\alpha$) and the intrinsic S/N at the highest frequency subband if the FRB was at the boresight of the beam ($S_{\rm t}$). 
The dashed lines mark the 2.5$\%$, 50$\%$ and the 97.5$\%$ credible intervals while the errors on the parameters are 2-$\sigma$ (68$\%$) errors. Both $\phi_{X}$ and $\phi_{Y}$ are in units of pixels with the origin at the bottom left corner and the resolution of 1 arcsecond/pixel such that the boresight of the beam is located at a coordinate of (100,100). \textbf{Right Panel:}GMOS-S $r$-band image with the point spread function of the coherent beam in which FRB~20201123A was discovered, overlaid on top (blue lines). The contours are at 0.0001$\%$, 0.4$\%$, 3$\%$, 18$\%$ and 99.9$\%$ levels. The 25$\%$ power level of the coherent beam is marked by the dashed white curve. One can see that the sidelobes of the beam are coincident with the primary beam of neighbouring coherent beams to rule them out as favourable locations (see text for more details) . The 99$\%$ confidence localisation region is shown in red and the most probably host galaxy for the FRB, \jname~is marked by the green ellipse.
}
\label{fig:loc} 
\end{figure*}

\noindent
%where $\alpha$ is the intrinsic spectral index, 
where $S_{\rm t}(\nu_i)$ is the \textit{true} integrated S/N in band $i$ (i.e. that would have been observed had the FRB occurred on boresight), $S_{\rm o}(\nu_i)$ is the observed integrated S/N in band $i$, $\pmb{\phi}$ is a two-dimensional vector that represents the directional offset of the FRB from the boresight of the beam and $G(\pmb{\phi}, \nu)$ is the beam response of the telescope. Assuming an ideal telescope receiver (i.e a flat response of the receiver across the entire band), the $S_{\rm t}$ for a given frequency $\nu$ would scale as,
\begin{equation}
     S_{\rm t}(\nu_i) = S_{\rm t}(\nu_{0})\left(\frac{\nu_i}{\nu_{0}}\right)^{\alpha},
\end{equation}
where $\nu_{0}$ is the highest frequency subband. The observed dynamic spectrum of FRB~20201123A can be approximated as a power-law and hence that assumption is valid. In order to perform the analysis, we split the data for FRB~20201123A into 4 frequency sub-bands and for each band we computed the S/N of the detected burst. $G$ was computed from the actual point spread function of the coherent beam where the FRB was detected. We used the \textsc{Mosaic}~\citep{chen2021} software to generate the PSFs for various subbands. Each PSF is 200x200 pixels wide with a resolution of 1 arc-second/pixel. The resulting $G$ and S/N values were fed into the Bayesian framework to compute the posterior probability of the location of the FRB as shown in Figure~\ref{fig:loc}. We used flat priors for $\alpha$, $\phi$ and $S_{\rm t}(\nu_{0})$. As one can see, the posterior distribution of $\phi_{X}$ and $\phi_{Y}$ is degenerate with multiple local maxima. This suggests that while we can characterize the spectrum of FRB~20201123A as a power-law, one cannot break the degeneracy between the position of the burst in the beam and the S/N of the burst if we assume that bursts with extremely high S/N ratios ($\geq$ 1000) are as likely as bursts with lower S/N ratios. Constraining the higher end of the prior on $S_{\rm t}$ will only bias the posterior to favour the boresight of the beam. If there are no constraints on the priors, the 2-D posterior is expected to follow the PSF of the CB which is exactly what is seen in the left panel of Figure~\ref{fig:loc}. Those regions can be ruled out by the fact that the FRB was not seen in any adjacent coherent beam which should have been the case if it were in any of those locations (see right panel of Figure~\ref{fig:loc}). Hence, for the purposes of this analysis, the FRB is equally likely to be anywhere in the CB. This is a limitation of this technique in the case of a single beam detection. However, if the FRB is detected in multiple beams, one can compute the joint posterior distribution of the location of the burst to get more precise constraints on the location using this technique.
%We do note that there are a few caveats here that can increase the 99$\%$ credible interval closer to the FWHM of the beam. For example. we have not considered the response of the bandpass in this analysis. We assume that to be flat across all frequency channels which certainly is not the case. However, we do note that typically, the bandpass response of MeerKAT can be approximated as a power-law and will only affect the estimate of the spectral index $\alpha$ without affecting the estimate of the location of the burst but will certainly affect the width of the posterior distribution.
Closely spaced beams where the beams overlap at a significant fraction of the maximum response will help constrain the location of the burst greatly in absence of detections in neighbouring beams. To overcome these limitations, we also decided to use another Bayesian framework to hone in on the location of the burst and in turn, identify the possible host galaxy for FRB~20201123A which we describe in the section below.

%Hence, we combine this analysis with another Bayesian framework \citep{path2021}
%to hone in on the location of the burst and in turn, identify the possible host galaxy for FRB~20201123A.

\subsubsection{Host Association}
% 2021-04-13 -> 3 images of 100 s each (not used)
% 2021-04-14 -> 20 images of 100 s each (19 used only)
% 2021-05-15 -> 16 images of 100 s each (all used)
% We used only the good quality subset, corresponding to 35 images
On 2021-04-14 and 2021-05-15 UT, 
we obtained a series of $35 \times 100$\,s $r$-band images of the field surrounding FRB~20201123A with the
Gemini Multi-Object Spectrograph \citep[GMOS;][]{GMOS} mounted on the Gemini-South telescope as part of program GS-2021A-Q-134.
Given the low Galactic latitude ($b \approx -10$\,deg), the field is crowded by stars, and thus we prioritized high image quality. These data were reduced with standard image processing
techniques 
%(e.g.\ bias subtraction, flat fielding, coaddition) 
using the DRAGONS software.\footnote{\url{https://dragons.readthedocs.io}} 
The final stacked $3\,500$\,s image has an effective PSF FWHM of $\approx 0.64$\arcsec, and it was astrometrically calibrated to match the Gaia DR2 catalog \citep{GaiaDR2, GaiaDR2_astro} with an astrometric accuracy of $0.2$\arcsec. %(relative astrometric accuracy of $XX\arcsec$). 
%The astrometric solution provided with the images was further refined using Gaia stars identified in the field.
%The images were then registered and combined and then flux calibrated with XXXX.

Figure~\ref{fig:201123} shows the $\approx 50$\arcsec\
diameter localisation of FRB\,201123 on the 
combined $r$-band image.
Because of the single-band detection, the localisation
is nearly uniform within its boundary.
In this region, one expects many tens of galaxies
which challenges the association of FRB\,201123
to its host galaxy.  
%The image reveals a further complication -- the presence of numerous stars along this low latitude ($b \approx -10$\,deg) sightline through our Galaxy.
The association is further complicated by the presence of numerous stars and a 
%Galactic reddening of $E(B-V) = 0.19$\,mag 
Galactic extinction of $A_r \approx 1.56$\,mag. However, the image reveals a single bright galaxy 
(hereafter \jname)
towards the north-west of the region. Galaxies with its apparent magnitude
($m_r = 15.97$\,mag; corrected for Galactic extinction 
and by the presence of an interloper star)
and half-light size ($\phi \approx 2''$)
are very rare and one is intended to
favor this system as the host on chance
considerations alone.

We proceeded to perform a probabilistic 
association to transient host 
\citep[PATH;][]{path2021} analysis for FRB~20201123A.
From the $r$-band image, we used the {\sc photutils} 
package \citep{photutils2021}
to detect sources within the localisation region.
Table~\ref{tab:frb201123} lists all of these detected sources
and their measured properties. Apparent magnitudes were calibrated using reference stars from the SkyMapper Southern Survey \citep{skymapper}. 
In order to assess whether an object is a galaxy, we run {\sc Source Extractor} \citep{bertin1996} and used the {\tt CLASS\_STAR} parameter as our star/galaxy classifier. In the following, we restrict to sources with 
%{\tt CLASS\_STAR}$<0.9$
a star/galaxy classifier value lower than 0.9
and assume the remainder are stars. Four of the sources detected by {\sc photutils} were not detected by {\sc Source Extractor} and thus these lack a star/galaxy classifier; in the following we will conservatively assume that these are all galaxies.
Last, we correct the apparent magnitudes
by $A_r \approx 1.56$\,mag for Galactic extinction.

\begin{table}
%\centering
\small
\caption{Sources within the localization region of FRB201123 and their properties.
$\phi$ is the half-light radius as defined by PATH and the $r$-band magnitudes $m_r$ are not corrected for Galactic extinction.
$P(O)$ and $P(O|x)$ are the prior and posterior probabilities,
respectively, for each galaxy candidate.}
\label{tab:frb201123}
\begin{tabular}{ccccccc}
\hline
RA & DEC & $\phi$ & $m_r$ & object & $P(O)$ & $P(O|x)$\\ 
(deg) & (deg) & ($\arcsec$) & (mag) & classifier & & \\ 
\hline
 
263.66814 & -50.76320& 0.30& 18.6& 0.99& 0.000 & 0.000\\ 
263.66655 & -50.76323& 0.18& 20.9& 0.98& 0.000 & 0.000\\ 
263.65963 & -50.76407& 1.75& 16.5& 0.03& 0.838 & 0.915\\ 
263.67147 & -50.76451& 0.24& 20.1& 0.98& 0.000 & 0.000\\ 
263.65821 & -50.76454& 0.20& 20.9& -- & 0.006 & 0.008\\ 
263.67429 & -50.76465& 0.22& 20.3& 0.98& 0.000 & 0.000\\ 
263.66691 & -50.76538& 0.19& 20.9& 0.99& 0.000 & 0.000\\ 
263.65972 & -50.76561& 0.15& 21.7& -- & 0.003 & 0.004\\ 
263.67249 & -50.76610& 0.34& 17.8& 0.99& 0.000 & 0.000\\ 
263.66119 & -50.76641& 0.21& 20.6& -- & 0.008 & 0.011\\ 
263.65584 & -50.76682& 0.26& 19.4& 0.98& 0.000 & 0.000\\ 
263.67210 & -50.76691& 0.23& 20.1& 0.98& 0.000 & 0.000\\ 
263.67667 & -50.76731& 0.15& 21.5& 0.98& 0.000 & 0.000\\ 
263.66322 & -50.76732& 0.63& 22.2& -- & 0.002 & 0.002\\ 
263.66439 & -50.76746& 0.19& 20.7& 0.98& 0.000 & 0.000\\ 
263.65874 & -50.76765& 0.29& 18.9& 0.99& 0.000 & 0.000\\ 
263.66397 & -50.76782& 0.29& 18.9& 0.98& 0.000 & 0.000\\ 
263.66740 & -50.76814& 0.52& 15.6& 0.96& 0.000 & 0.000\\ 
263.66128 & -50.76787& 0.24& 20.0& 0.88& 0.014 & 0.019\\ 
263.65673 & -50.76838& 0.35& 17.7& 0.98& 0.000 & 0.000\\ 
263.67182 & -50.76831& 0.24& 19.9& 0.98& 0.000 & 0.000\\ 
263.67399 & -50.76835& 0.24& 19.9& 0.98& 0.000 & 0.000\\ 
263.66306 & -50.76873& 0.38& 16.6& 1.00& 0.000 & 0.000\\ 
263.65996 & -50.76859& 0.31& 18.6& 0.98& 0.000 & 0.000\\ 
263.67438 & -50.76912& 0.29& 18.8& 0.99& 0.000 & 0.000\\ 
263.65887 & -50.76920& 0.30& 18.7& 0.98& 0.000 & 0.000\\ 
263.67301 & -50.76945& 0.29& 19.1& 0.98& 0.000 & 0.000\\ 
263.67363 & -50.76963& 0.16& 21.2& 0.98& 0.000 & 0.000\\ 
263.66927 & -50.76982& 0.34& 17.7& 0.95& 0.000 & 0.000\\ 
263.66560 & -50.77039& 0.28& 19.3& 0.84& 0.030 & 0.041\\ 
263.66719 & -50.77070& 0.51& 15.8& 0.99& 0.000 & 0.000\\ 
263.65809 & -50.77129& 0.28& 19.2& 0.98& 0.000 & 0.000\\ 
263.66104 & -50.77129& 0.14& 21.7& 0.98& 0.000 & 0.000\\ 
263.67622 & -50.77218& 0.18& 20.8& 0.98& 0.000 & 0.000\\ 
263.67074 & -50.77272& 0.28& 19.7& 0.98& 0.000 & 0.000\\ 
263.67280 & -50.77278& 0.23& 20.0& 0.98& 0.000 & 0.000\\ 
263.65951 & -50.76679& 0.23& 20.7& 0.98& 0.000 & 0.000\\ 
263.65988 & -50.76690& 0.20& 20.9& 0.98& 0.000 & 0.000\\ 
\hline 
\end{tabular}
\end{table}

In addition to the localisation and candidate
galaxies, 
one must also adopt a set of priors  
to perform the PATH analysis.
We follow the preferred assumptions
of \cite{path2021}, i.e.\ their inverse prior and 
exponential offset function 
with $\theta_{\rm max} = 6\phi$.
For the unseen prior $P(U)$ we assume a value
of $10$\% based on the angular sky coverage
of the stars in the field. The last two columns of Table~\ref{tab:frb201123}
list the prior probabilities $P(O)$
and posterior probabilities $P(O|x)$ for
each of the galaxy candidates.
The PATH results clearly favor \jname\ as the host
of FRB~20201123A with a posterior probability
$P(O|x) = 0.92$.  We caution, however, that given the
uniform localisation of this FRB, this result is
primarily driven by its bright flux and large
angular size.
Nevertheless, we proceed with relatively high
confidence in this association (e.g.\ its posterior
probability exceeds that of every other candidate by
an order-of-magnitude).

%We performed an additional test with PATH to account 
%for the high incidence of Galactic stars 
%along this sightline.
%Specifically,
%we considered the possibility of one or more
%relatively bright galaxies 
%obscured by stars in the field and re-estimated
%the posteriors.  This treatment is similar to 
%adopting a non-zero $P(U)$ value as above.
%For an assumed $5$ galaxies each 
%with $m_r = 21$\,mag, a random angular size
%between $1\arcsec$ and $2\arcsec$ and keeping the 
%unseen prior at $10\%$, we find the posterior probability 
%for \jname\ is $0.992$.
%Deeper imaging yielding yet fainter and
%smaller galaxies will have a minor effect. 

\begin{figure*}
 \centering
\includegraphics[width=0.9\textwidth]{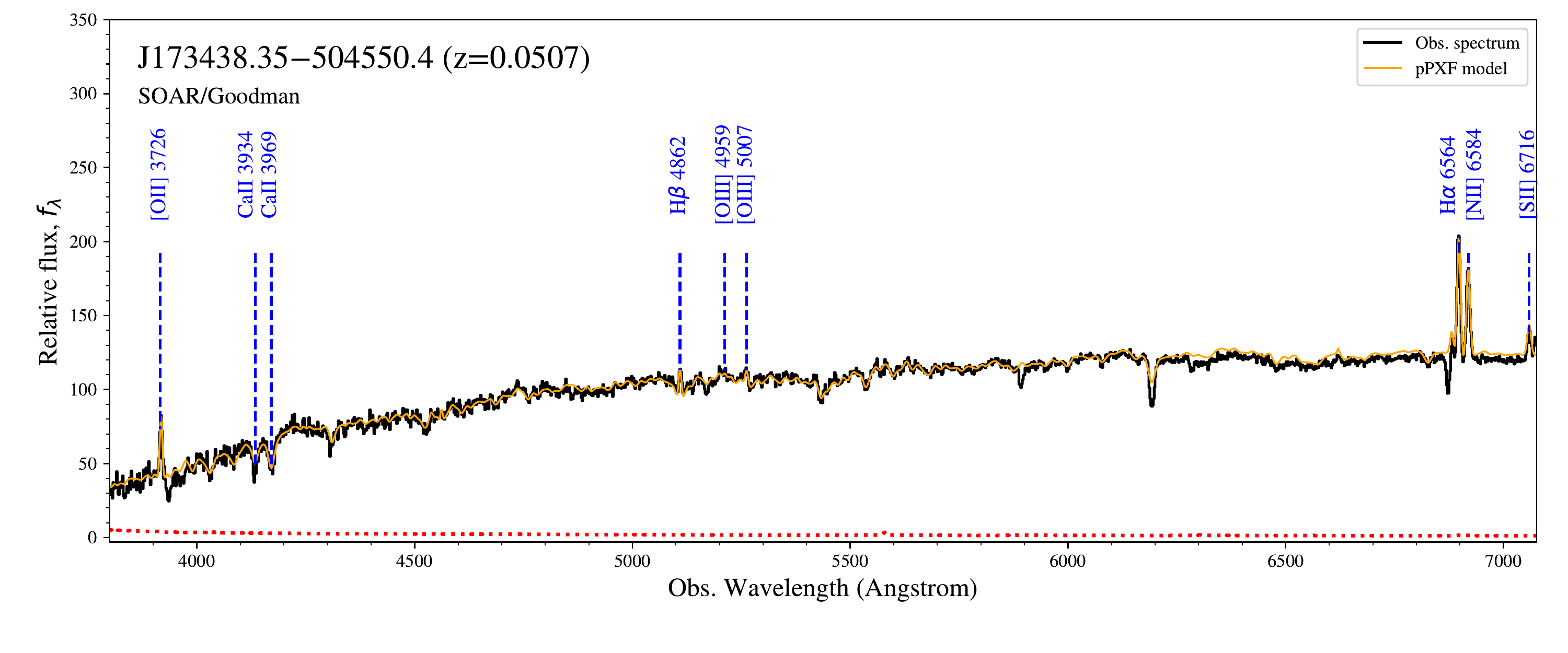}

\caption{SOAR/Goodman optical spectrum of \jname, the putative galaxy host of FRB~201123. The observed spectrum is shown by the black histogram and its uncertainty is shown by the red dashed line. Some spectral features at a common redshift of $z=0.0507$ have been highlighted in blue. Our adopted pPXF model is shown as a solid orange line. See Section~\ref{sec:host} for further details.
%We  at $\lambda < 5500$\,\AA, with spectral features at a common redshift of $z=0.0507$ marked. \textbf{Right Panel:} A zoom-in around the \ha\ and \nii\ emission lines. \nii\ $\lambda$6548 is affected by Telluric absorption.}  %Unfortunately, these are severely affected by Telluric absorption and fluxing errors which preclude their quantitative analysis.
}
\label{fig:spectrum} 
\end{figure*}

\subsubsection{Putative Host Analysis}
\label{sec:host}

Adopting \jname\ as the putative host galaxy
for FRB~20201123A, we now proceed to measure its
properties and compare these to other, secure
FRB hosts \citep{heintz+2020,bhandari+2021}.
Figure~\ref{fig:spectrum} shows a spectrum of
\jname\ obtained on 2021-03-24 UT with the Goodman spectrograph \citep{goodman}
on the SOAR telescope as part of program SOAR2021A-010.
The instrument was configured with the 400\_SYZY
grating, a 1.0\arcsec\ long slit, and $2 \times 2$~binning.
These data were reduced with the PypeIt data
reduction pipeline \citep{pypeit} and flux
calibrated with a spectrophotmetric standard
requiring the $r$-band apparent magnitude match that
of the galaxy (to crudely correct for slit losses).
For the figure and subsequent analysis, we have
corrected the data for Galactic extinction 
assuming a reddening $E(B-V) = 0.19$\,mag
and the \cite{ccm89} extinction law. % https://ui.adsabs.harvard.edu/abs/1989ApJ...345..245C/abstract
The galaxy exhibits strong nebular emission lines
at a common redshift $z=0.0507$ including
\oii, \hb, \oiii, \ha\ and \nii.

We performed a stellar population fit to the data
at wavelengths $\lambda_{\rm obs} = 3700-7075$\AA\ 
using the pPXF software package \citep{ppxf}.
For the analysis, we followed the assumptions of
previous works \citep[e.g.][]{heintz+2020}.
The results provide an estimate of the stellar
mass ($M* \approx 10^{\mstar} M_\odot$) and the emission-line
flux of H$_\alpha$ 
($f_{\rm H \alpha} = 4.2 \times 10^{-15}$\,ergs~cm$^{-2}$~s$^{-1}$) with
the latter yielding an inferred star formation
rate of SFR~$\approx \sfr M_\odot \, \rm yr^{-1}$
after correcting for internal extinction with
$A_V \approx 0.7$\,mag based on the 
H$_\alpha$/H$_\beta$ ratio.
Uncertainties in $M*$ and SFR are dominated by
systematics which include the assumed IMF,
slit loss, and dust extinction (both Galactic
and internal).  
We estimate 0.3~dex errors for each quantity.
The $M*$ and SFR values
of \jname\ place it well within the locus of
measured values of highly secure FRB host
galaxies \citep{bhandari2021}.

\section{Fluence limits}
Apart from FRB~20201123A, we do not have accurate localisation information on the newly discovered FRBs. This makes calculation of a fluence for the FRBs quite challenging. Hence, we estimate the lower limit on the fluences of the bursts if we assume that the FRB~20200413A and FRB~20200915A are close to the edge of the CB tiling pattern and FRB~20201123A is at the boresight of the CB in which it was discovered. Using the radiometer equation~\citep{dewey1985}, for an FRB with a signal-to-noise ratio $\rm S/N$, the fluence,
\begin{equation}
    \mathcal{F} = \frac{\rm S/N~T_{sys} \sqrt{W_{meas}}}{G\sqrt{n_{p}\Delta \nu}}\times 1000\, \rm Jy~ms,
\end{equation}
where, $\rm T_{sys}$ is the system temperature, $G$ is gain of the telescope in K~Jy$^{-1}$, $\rm W_{meas}$ is the measured width of the FRB in seconds (after de-dispersion and after compensating for any other smearing effects), $n_{p}$ is the number of polarizations summed and $\Delta \nu$ is the bandwidth of the receiver in Hz.
Instrumental parameters for MeerKAT from Bailes et al. (2020) are $n_{p}$=2, $\Delta \nu$= 856 MHz. We use a $G$=0.175 K/Jy for \frbA~and 0.3~K~Jy$^{-1}$ for \frbB~after taking into account the reduction in the gain of the IB telescope due to the offset from the boresight. This is because the power level at which the CB tiling ends is different for the two FRBs. We assume a gain of 1.75~K~Jy$^{-1}$ for the CB of FRB~20201123A (modified from 2.8 K/Jy since only 40 out of 64 MeerKAT dishes were used for the
observations). The telescope’s system temperature $\rm T_{sys}$ is given by
$\rm T_{sys}$ = $\rm T_{rec}$ + $\rm T_{sky}$, with $\rm T_{sys}$ = 18 K the receiver temperature for the 1.4~GHz observations~\citep{ridolfi2021}. We assume the total sky temperature to be 23~K for all FRBs assuming a mean of 5~K contribution due to the sky. Using these values, we obtain a lower limit on the fluence, $\mathcal{F}\gtrsim$3.1~Jy~ms and 3~Jy~ms for \frbA~and \frbB~respectively. For FRB~20201123A, we also correct the S/N to account for the reduction in the gain of the CB due to its offset from the boresight of the primary beam (18.072 arc-minutes). If we assume that FRB~20201123A originates from the host galaxy \jname, we can compute the true fluence of the burst. In order to do that, we obtain the true gain of the CB at the optical centre of \jname~by simulating the point spread function of the CB using \textsc{mosaic}~\citep{chen2021}\footnote{\url{https://github.com/wchenastro/Mosaic}}. We estimate $G$ of 1.4~K~Jy$^{-1}$ and using this new value, we obtain $\mathcal{F}\gtrsim$1.4~Jy~ms. The final limits are presented in Table~\ref{tab:params}. Given the redshift of \jname, we estimate the lower limit on the energy of the burst in the MeerKAT band of 8.4$\times$10$^{34}$~J. This luminosity is consistent with the general population of FRBs~\citep{luo2018}.

\section{Discussion}
\label{s:discussion}
\subsection{Limits on repeating bursts}
Since MeerTRAP is a commensal survey instrument, there are always opportunities to follow-up newly discovered FRBs as the field is usually observed multiple times by the LSPs. This means that we were able to obtain follow-up of the fields to look for repeat bursts for the three FRBs reported in this paper. FRB~20201123A shows a post-cursor about 200~ms after the initial bright burst suggesting that it may be a repeater. Such post-cursor bursts have been observed in other repeating FRBs like FRB~20121102A~\citep{zhang2018, Cruces2021} and could be a vital diagnostic to identify the nature of FRB~20201123A. So far, MeerTRAP has observed the same field for a total of 25.4 hours. No other burst was detected at the same DM at the S/N threshold of 8.0. If we assume that FRB~20201123A is a repeating source and the burst arrival times are Poisson distributed~\citep{Cruces2021}, one can obtain a 95$\%$ confidence limit on the repeat rate for the effective sensitivity of the survey in that direction (assuming that it remains unchanged in that direction for successive observations). From~\cite{gehrels1986}, the 95$\%$ confidence level upper limit on the repeat rate is given by
\begin{equation}
    \mathcal{R}_{ul} = \frac{4.744}{t_{\rm obs}},
\end{equation}
where $t_{\rm obs}$ is the total observation time. This gives a limit on $\mathcal{R}_{ul}$ of 0.18 bursts per hour. For the other two FRBs, the total time on sky is too low (< 2 hours) to put any tight constraint on the repeat rate. We note that the gap between detection of the first event and the start of the follow-up observation is important to find repeat bursts~\citep[see][for more details]{caleb2019} but since MeerTRAP does not have control over when the same field is observed, we can only provide rough limits on the repeat rate of these FRBs.

\begin{figure}
 \centering
\includegraphics[width=3.7 in]{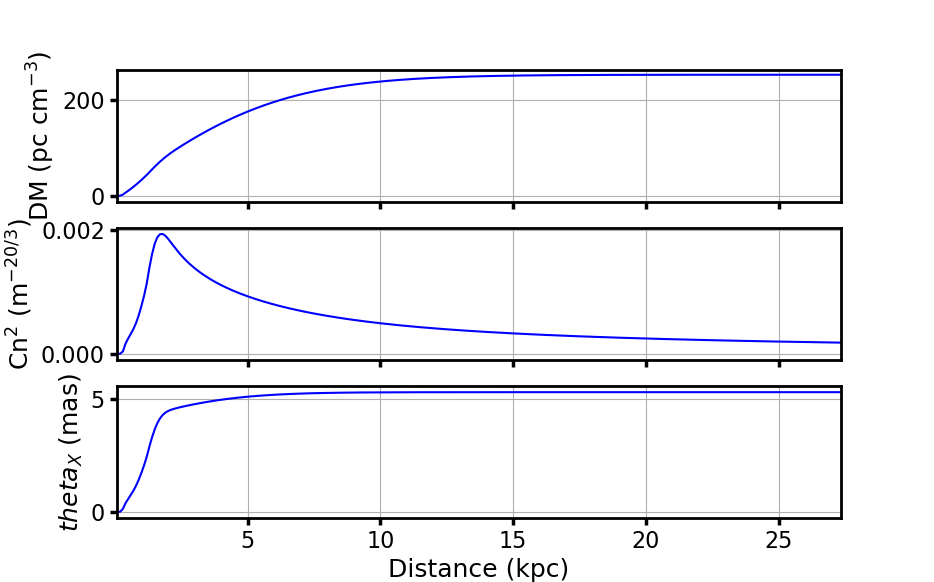}
\caption{The expected galactic DM contribution (top panel), strength of scattering (middle panel) and angular broadening of extragalactic sources (bottom panel) as a function of distance from the Earth towards \frbC. The simulations were performed using \textsc{ne2001} model (see text for more details).}
\label{fig:ne2001} 
\end{figure}

\subsection{Scattering in \frbC}
The spectral index of the scattering in \frbC~($\alpha$) (the left panel of figure~\ref{fig:scatter}) is $-$4.2
which is consistent within 1-$\sigma$ errobars to the $\nu^{-4.4}$ relationship that is expected for a turbulent
medium that follows a Kolmogorov spectrum~\citep[e.g][]{handbook}. In order to determine the predicted scattering
contribution from the Milky Way, we looked at the expected values from the electron density models of our Galaxy
along this line of sight~\citep{ne2001,ymw16}.  For an FRB at a distance $D$ and a scattering screen distance, $d_{\rm screen}$, $\tau \simeq \frac{\theta~d^{2}_{\rm screen}}{8ln(2)c}$ where $\theta$ is the extragalactic
angular broadening and $c$ is the speed of light~\citep[see][and the references therein]{main2021}. We took 100
randomly sampled Right Ascensions and Declinations within a 30"x30" circle around the centre of the CB in which
FRB~20201123A was detected and computed the total Scattering Measure and the DM contribution of our Galaxy as a
function of distance using the electron density models. Figure~\ref{fig:ne2001} shows the values of $C_{n}^{2}$ (the
measure of scattering strength) and the DM contribution and one can clearly see that the maximum contribution to
scattering comes from a putative screen at a distance of 1.8~kpc. Using 
$\theta \sim$5.314~mas gives $\tau \sim 22 \,  \mu$s at 1~GHz. The \textsc{ymw16} model~\citep{ymw16} reports an
expected scattering timescale of 150~$\mu$s which is still two orders of magnitude lower than the measured
$\tau\sim$10~ms at 1~GHz, suggesting that the scattering cannot originate from our Galaxy. This means that the scattering in this FRB might originate from an extremely turbulent medium close to the FRB itself as is inferred from the current FRB population~\citep[see][]{chawla2021}. 

Another possibility that we considered was
that the majority of the scattering originates from an intersecting halo of a foreground galaxy~\citep[see][for more details]{prochaska2019}. While it is expected that the gas in halos of galaxies is not dense, the geometric boost
under the thin screen scattering model may render an intervening halo to dominate the scattering comparable to turbulent
environments in the host or our Galaxy if it is between the host galaxy and us. In order to quantify how
likely it is that the sightline is intersected by foreground halos, we follow the same technique as presented
in~\cite{prochaska2019}. We use the Aemulus Halo Mass function~\citep{mcclintock2019} to generate halos of masses
between 2$\times$10$^{10}$~$\rm M_{\odot}$ and 10$^{16}$~$\rm M_{\odot}$. Then, we compute the average number of
halos in the comoving volume enclosed at the redshift of the host galaxy of FRB~20201123A (z=0.0507), $N(z)$=0.227. Then, assuming the impact parameter of the FRB line of sight to be comparable to the virial radius of the intervening galaxy and assuming that the halos are Poisson distributed in the given comoving volume, the probability that the sightline is
intersected by $k$ halos,
\begin{equation}
 P(k|N(z)) = \frac{N^{k}~e^{-N}}{k!}.
\end{equation}
Hence the probability of intersecting at least one halo, $P(k\gtrsim1|N(z))$ is 1$-e^{-N}$ which is $\sim$21$\%$. The value of $P(k\gtrsim1|N(z)$ is sensitive to the value of $N(z)$ which in turn depends on the lowest halo mass assumed in the Halo Mass function. If we assume that the lower limit on the halo mass comes from the halo of a Milky Way like galaxy (10$^{12}$~$\rm M_{\odot}$), we get $N(z)$ of 0.099 which  reduces $P(k\gtrsim1|N(z))$ by a factor of 2. Regardless, this is a non-negligible probability and allows that the scattering could originate from multiple intersecting halos of foreground galaxies.

\begin{figure*}
 \centering
\includegraphics[width=3.5 in]{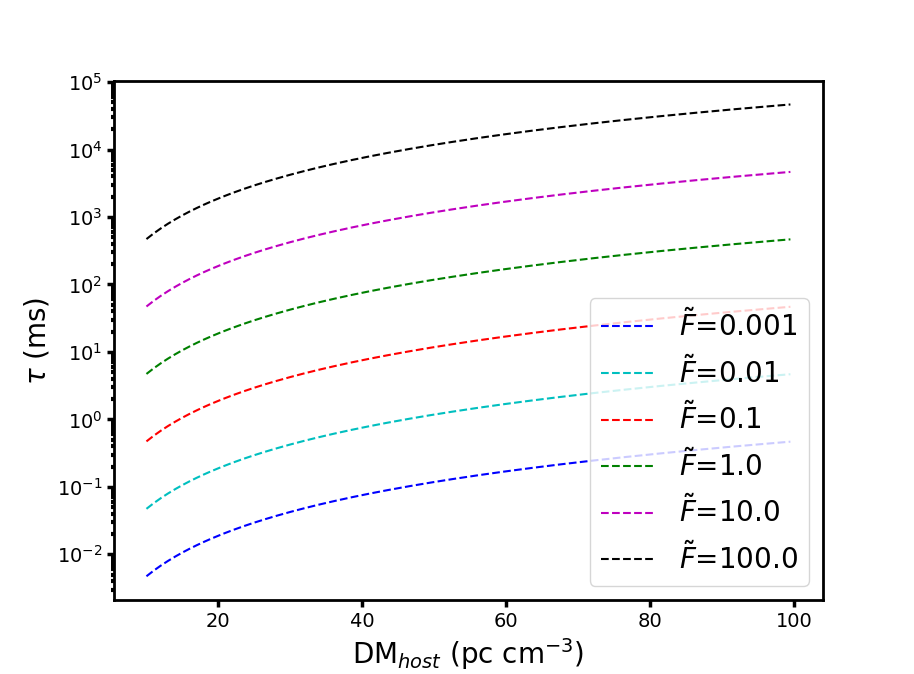}
\includegraphics[width=3.5 in]{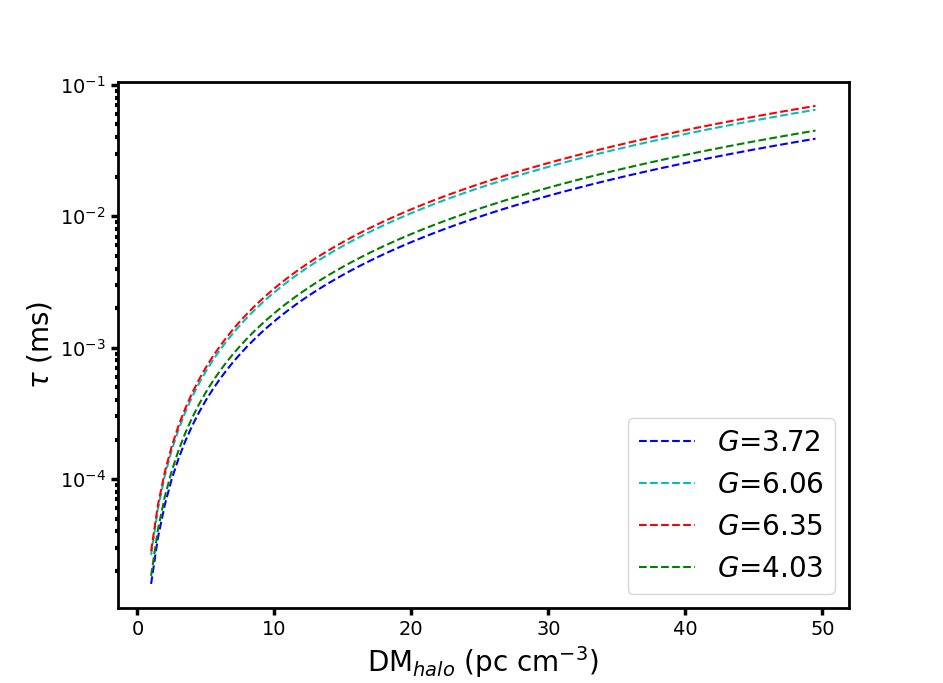}
\caption{\textbf{Left Panel}: Scattering time ($\tau$) at 1~GHz as a function of host DM contribution for various values of \protect{$\tilde{F}$} that signifies turbulence for \frbC. \textbf{Right Panel}: \protect{$\tau$} at 1~GHz as a function of DM from the halo of an intervening galaxy for values of $G$ (computed from a range of redshifts between 0 and 0.0507) that signifies the geometric boosting to scattering. The analysis is done for a lens with a size of 0.03~Mpc}
\label{fig:hosthaloscatter} 
\end{figure*}

In order to distinguish between the two possibilities we predict the scattering that would be induced if it was originating in the host or originating in an intervening galaxy. Based on previous work by~\cite{ocker2021}, the expected scattering time,
\begin{equation}
\tau (\nu, {\rm DM}, z) \simeq 48.03~\frac{A_{\tau}\tilde{F}G~{\rm {\rm DM_{l}^{2}}}}{(1+z_{l})^{3}~\nu^{4}}\, \mu s,    
\end{equation}
where $A_{\tau}$ is a factor close to unity, $\tilde{F}$ is factor that characterizes the turbulence in the scattering medium, $G$ is the geometric boosting  to scattering due to the distance between the scattering screen, the source and the observer ($G \sim\, d_{lo}d_{sl}/L~d_{so}$) where $d_{sl}$, $d_{so}$ and $d_{lo}$ are the angular diameter distances of, source to lens, source to observer and lens to observer respectively. $\rm DM_{l}$ is the DM contribution from the scattering medium, $z_{l}$ is the redshift of the scattering medium and $\nu$ is the observing frequency. If the scattering originates in the host galaxy, we expect $\tilde{F}$ to dominate the total scattering time while $G\sim 1$. On the other hand, if the scattering is dominated by the intervening galaxy halo, we expect $G$ to be large ($G \simeq 2\delta d/L$) for $\delta d \ggg L$ where $\delta d$ is the distance from the lens to the source or the observer. In this case, we expect little to no turbulence from the diffuse gas as shown from previous studies~\citep[see][for more details]{prochaska2019}. Here, we compute the scattering for a foreground galaxy that lies at a distance ranging from 25$\%$ to 75$\%$ of the redshift of the host. For the host galaxy scattering, we assume $G\sim$1 and compute scattering for various values of $\tilde{F}$. For scattering from an intervening halo, we used the smallest value of $\tilde{F}\sim 0.0001$ measured for pulsars in our Galaxy~\citep{ocker2021} and computed scattering for various values of $G$ derived from the range of distances of the intervening halo. We also evolve $\tilde{F}$ with the Star-Formation rate using Eq.~21 in~\cite{ocker2022}. We note that the value of $\tilde{F}$ used is very conservative as $\tilde{F}$ is expected to be even smaller but pulsars are unable to probe this diffuse gas. Figure~\ref{fig:hosthaloscatter} shows the expected scattering for both scenarios. It is clear that the foreground galaxies cannot account for the scattering seen in \frbC~(7.5~ms at 1~GHz) even for all possible values of $G$ within the co-moving value enclosed by the host galaxy of \frbC~while one can obtain the expected $\tau$ at 1~GHz from the host galaxy itself fairly easily. Hence, we conclude that the scattering in this FRB originates from the host galaxy.

\subsection{Host galaxy of FRB~20201123A}
Along with the time-domain detections, we identified \jname~as a potential host galaxy for FRB~20201123A using robust statistical treatment given the relatively small localisation error region. At face value, the low redshift of \jname\ appears at odds with the large dispersion measure for FRB~20201123A ($\mdmfrb \approx 434 \, \mdmunits$). Our Galaxy, however, contributes $\mdmism \approx 200 \, \mdmunits$  (NE2001 {\rm gives}\, 229\,\dmunits\ and YMW16 gives 162~\dmunits) from its interstellar medium and
a presumed $\mdmhalo \sim 50 \, \mdmunits$ from its halo \citep{xyz19}. % https://ui.adsabs.harvard.edu/abs/2019MNRAS.485..648P/abstract 
This leaves $\approx 180 \, \mdmunits$ for the cosmos (\dmcosmic) and the host (\dmhost). At $z=0.05$, the average cosmic contribution is $\langle \mdmcosmic \rangle\sim42\, \mdmunits$ \citep{macquart2020} but the intrinsic scatter in this quantity is predicted to be large. Adopting the best-fit model to the Macquart
relation by \cite{macquart2020}, the 95\%\ confidence interval is $\mdmcosmic = [15,125] \, \mdmunits$. Allowing for the maximum value of this interval
(which would imply a significant foreground galaxy halo), we recover a minimum host 
contribution of ${\rm DM}_{\rm host, min} \approx 60~\rm pc~cm^{-3}$. This is consistent with estimates for host galaxy contributions from theoretical and empirical treatments \citep{xyz19,james2021}. For a true \dmcosmic\ value of this sightline closer to (or below) the mean, the host contribution would exceed $100 \, \mdmunits$. Such values are inferred for other FRB hosts \citep[e.g. FRB\,20121102A;][]{tendulkar2017}. In conclusion, we find no strong evidence to rule out the association with \jname\ based on its redshift and \dmfrb. The significant host contribution to the DM combined with the scattering in \frbC~ possibly originating in the host shows that it shares similarities with other highly active repeating FRBs like FRB~20121102A and FRB~20190520A and potentially resides in a turbulent and dense environment within the host.

\subsection{Speckled emission of FRB~20200915A}
As mentioned previously, we see speckled emission in the dynamic spectrum of FRB~20200915A whose frequency evolution may not with interstellar scintillation from our Galaxy. A more robust method of confirming this conjecture would be to fit the scintles analytically to show how the width varies with frequency. In the case of \frbB, the scintles are not bright enough or do not conform to a specific shape such that they could be fit by an analytical model. Hence, the disagreement with Galactic scintillation can only be marginally validated visually in this case. Another possibility is that the speckled behaviour may arise from plasma lensing of FRB emission in the vicinity of the progenitor giving rise to caustics that are observed in the dynamic spectrum. The caustics are chromatic in nature and can manifest themselves as magnified islands of emission in time and observed frequency space~\citep[see][for more details]{Cordes2017,main2018}. Within a perfect-lens approximation, the frequency width of a caustic due to magnification from a 1D elliptical lens,
\begin{equation}
    \frac{\Delta\nu}{\nu} \propto \frac{R_{1}^{2}}{\Delta x^{2}},
    \label{eq:lens}
\end{equation}
where $R_{1}$ is the equivalent radius ($R_{1}$ = $R_{\rm fr}$/$\sqrt{\pi}$ where $R_{\rm fr}$ is the Fresnel scale) of the lens in the circular approximation and $\Delta x$ is the semi-major axis of the lens. One caveat here is that the relative velocity between the source and lens will cause the waves to interfere and produce complex patterns in the frequency space that can manifest itself as islands of emission drifting in time and frequency. Since we do not see any drift in time as seen for other repeating FRBs~\citep{hessels2019}, we can only put an upper limit on the velocity of the plasma lens in the $x$-direction with respect to the FRB. The relative velocity,

\begin{equation}
    v \geq {\rm 0.5}\frac{R_{1}}{\Delta t}~\left(\frac{\Delta \nu}{\nu}\right)^{0.5},
    \label{eq:v}
\end{equation}
where the delay in obtaining a new caustic due to the relative motion is,
\begin{equation}
    \Delta t \geq 0.7\frac{R_{1}^{2}}{v\Delta x}.
    \label{eq:t}
\end{equation}
In our case, $\Delta t$ is constrained by the maximum dispersion smearing in our band which corresponds to 3.4~ms for FRB~20200915A. Assuming the size of the filament close to the FRB of $\sim$1~AU, we get $R_{1}\sim$100~km. Using these values in Eq.~\ref{eq:v} gives a lower limit on $v$ of 4.6$\times$10$^{2}$~km~s$^{-1}$. Regardless of which model we consider, they cannot explain the decrease in the spectral extent of the emission with increasing frequency. Thus, we conclude that the speckled emission seen in \frbB~ might be intrinsic to the FRB. Such speckled emission as also been observed in a number of FRBs discovered by ASKAP~\citep{shannon2018} and could point to a common emission characteristic within the population. However, as mentioned previously, we cannot rule out weak scintillation arising from the halo of an intervening galaxy as confirming the same would require data at much higher time resolution.

\section{Summary}
\label{s:summary}

In summary, we present the first three discoveries of FRBs from the
MeerTRAP project. FRB~20200915A shows speckled emission structure in the
dynamic spectrum that is reminiscent of a few FRBs discovered by ASKAP.
We show that the decorrelation bandwidth of the scintles of
FRB~20200915A does not seem to follow the typical $\nu^{4}$
frequency relation and also cannot be explained by plasma lensing in the host galaxy, suggesting that the speckled nature of the burst might be intrinsic in nature. FRB~20201123A shows clear evidence for scattering. We investigated the origin of scattering in FRB~20201123A
and find that the scattering cannot originate from our Galaxy or the halo of an intervening galaxy and
is most likely dominated by some turbulent material in close proximity
to the source as expected from recent simulations by~\cite{chawla2021}.
As FRB~20201123A was detected only in a single coherent beam, we were
able to put tight constraints on its location. Using the non-detection
in adjacent beams and a Bayesian framework called
\textsc{path}, we were able to hone in on the most probably host galaxy
for FRB~20201123A. Assuming \jname\, to be the host of FRB~20201123A,
the DM contribution from the host is still consistent with what one
would expect for the host galaxy contribution to the DM which can exceed
100~pc~cm$^{-3}$ for the smallest contribution from the IGM based on the
Macquart relation. This combined with the scattering suggests
some similarities between the environment of~\frbC~and few of the
prolific repeating FRBs. None of the bursts were seen to repeat
although a faint post-cursor was seen 200~ms after the main burst of
\frbC.

\section*{Acknowledgements}

K.M.R., B.W.S., M.C., F.J., M.S., T.B., L.D., S.S., M.M. and V.M. thank the MeerKAT Large Survey Project teams for allowing MeerTRAP to observe commensally. K.M.R., B.W.S., M.C., F.J., M.S., T.B., L.D., S.S., M.M. and V.M. acknowledge funding from the European Research Council (ERC) under the European Union's Horizon 2020 research and innovation programme (grant agreement No. 694745). The authors also acknowledge the usage of TRAPUM infrastructure funded and installed by the Max-Planck-Institut für Radioastronomie and the Max-Planck-Gesellschaft. The authors would like to thank Christopher Williams and Aris Karastergiou for invaluable help during the commissioning of the MeerTRAP system. K.M.R. acknowledges support from the Vici research program 'ARGO' with project number 639.043.815, financed by the Dutch Research Council (NWO). M.C. acknowledges support of an Australian Research Council Discovery Early Career Research Award (project number DE220100819) funded by the Australian Government and the Australian Research Council Centre of Excellence for All Sky Astrophysics in 3 Dimensions (ASTRO 3D), through project number CE170100013. J.K.K acknowledges support from the Swiss National Science Foundation under grant 185692. 
Authors W.F., C.K., J.X.P., and N.T. as members 
of the Fast and Fortunate for FRB
Follow-up team, acknowledge support from 
NSF grants AST-1911140 and AST-1910471.
N.T. and C.N. acknowledge support by FONDECYT grant 11191217. The authors would like to thank Rene Breton for invaluable discussions regarding the localisation of FRB~20201123A using the Bayesian Framework. The MeerKAT telescope is operated by the South African Radio Astronomy Observatory (SARAO), which is a facility of the National Research Foundation, an agency of the Department of Science and Innovation. The authors would also like to thank the South African Radio Astronomy Observatory staff for all the help during these observations. Based on observations obtained at the international Gemini Observatory, a program of NSF’s NOIRLab, which is managed by the Association of Universities for Research in Astronomy (AURA) under a cooperative agreement with the National Science Foundation on behalf of the Gemini Observatory partnership: the National Science Foundation (United States), National Research Council (Canada), Agencia Nacional de Investigaci\'{o}n y Desarrollo (Chile), Ministerio de Ciencia, Tecnolog\'{i}a e Innovaci\'{o}n (Argentina), Minist\'{e}rio da Ci\^{e}ncia, Tecnologia, Inova\c{c}\~{o}es e Comunica\c{c}\~{o}es (Brazil), and Korea Astronomy and Space Science Institute (Republic of Korea). The Gemini data were obtained from program GS-2021A-Q-134, and were processed using
the DRAGONS (Data Reduction for Astronomy from Gemini Observatory North and South) package.
The MALS data were processed using the MALS computing facility at IUCAA (https://mals.iucaa.in).

%%%%%%%%%%%%%%%%%%%%%%%%%%%%%%%%%%%%%%%%%%%%%%%%%%
\section*{Data Availability}
The data will be made available upon reasonable request to the authors.
The galaxy observations of \jname\ are available
at https://github.com/FRBs/FRB.



\bibliographystyle{mnras}
\bibliography{main} % if your bibtex file is called example.bib

%\begin{thebibliography}{99}
%\end{thebibliography}

%%%%%%%%%%%%%%%%%%%%%%%%%%%%%%%%%%%%%%%%%%%%%%%%%%

%%%%%%%%%%%%%%%%% APPENDICES %%%%%%%%%%%%%%%%%%%%%

%%%%%%%%%%%%%%%%%%%%%%%%%%%%%%%%%%%%%%%%%%%%%%%%%%

% Don't change these lines
\bsp	% typesetting comment
\label{lastpage}
\end{document}